\documentclass[a4paper, 11pt]{article}
\usepackage{graphicx}
\usepackage{epsf,amsmath,bbold,amsfonts,stmaryrd}

\usepackage[utf8]{inputenc}
\usepackage{mathrsfs}
\usepackage{appendix}
\usepackage{amssymb}
\usepackage{float}
\usepackage{color}
\usepackage{cite}
\usepackage{hyperref}
\hypersetup{pageanchor=false}
\usepackage{indentfirst}
\usepackage{url}
\usepackage{float}
\usepackage{caption}
\usepackage[numbers,square,comma,sort&compress,merge]{natbib}

\usepackage{subfigure}

\def\nn{\nonumber}

\def\be{\begin{equation}}
\def\ee{\end{equation}}

\def\bea{\begin{eqnarray}}
\def\eea{\end{eqnarray}}

\def\ba{\begin{array}}
\def\ea{\end{array}}

\def\bc{\begin{center}}
\def\ec{\end{center}}

\def\bl{\begin{flushleft}}
\def\el{\end{flushleft}}

\def\br{\begin{flushright}}
\def\er{\end{flushright}}

\def\bi{\begin{itemize}}
\def\ei{\end{itemize}}

\def\bt{\begin{tabular}}
\def\et{\end{tabular}}

\newcommand{\al}{\alpha_{\text{fov}}}
\newcommand{\ds}{ds^2_{\text{eff}}}

\numberwithin{equation}{section}

\begin{document}
\title{\textbf{QED Effect on Black Hole Shadow}}
\author{
Zezhou Hu$^{1}$, Zhen Zhong$^{2}$, Peng-Cheng Li$^{1,3}$, Minyong Guo$^{3*}$ and Bin Chen$^{1,3,4}$}
\date{}
\maketitle

\vspace{-10mm}

\begin{center}
{\it
$^1$Department of Physics and State Key Laboratory of Nuclear
Physics and Technology, Peking University, No.5 Yiheyuan Rd, Beijing
100871, P.R. China\\\vspace{1mm}

$^2$Department of Physics, Beijing Normal University, Beijing 100875, P. R. China\\\vspace{1mm}

$^3$Center for High Energy Physics, Peking University,
No.5 Yiheyuan Rd, Beijing 100871, P. R. China\\\vspace{1mm}

$^4$Collaborative Innovation Center of Quantum Matter, No.5 Yiheyuan Rd, Beijing 100871, P. R.
China

}
\end{center}

\vspace{8mm}

\begin{abstract}
In this work, taking the QED effect into account, we investigate the shadows of the static black hole with magnetic monopoles and neutral black holes in magnetic fields through the numerical backward ray-tracing method. For a static black holes with magnetic monopole, we obtain the relation between the shadow radius and the coupling constant. For neutral black holes in the uniform magnetic fields, we find that the shadow curves deviate very small from the ellipses for equatorial observer, and  we read the linear relation between the eccentricity and the coupling constant. For $\theta_o\neq\pi/2$, we find that the shadow curves can be well approximated by ellipses in most cases, except the case that  the magnetic field is very strong and the observer sits around the angle $\theta_o=\pi/4$ or $3\pi/4$. Moreover we extend our investigation to a neutral static black hole surrounded with a current loop.

\end{abstract}

\vfill{\footnotesize $^*$Corresponding author: minyongguo@pku.edu.cn.}

\maketitle

\newpage

\section{Introduction}

Since the first  image of the supermassive black hole in the center of M87$^\ast$ galaxy taken by the Event Horizon Telescope (EHT) \cite{Akiyama:2019cqa, Akiyama:2019brx,Akiyama:2019sww,Akiyama:2019bqs,Akiyama:2019fyp,Akiyama:2019eap}, the study of black hole physics has entered a new era. The image of black hole not just provides the direct evidence of the existence of black holes, but also encodes a lot of information about the black hole and its surroundings. It opens a new window to investigate various problems in astrophysics and gravity theories.  Thus, how to extract the information from pictures of black holes or in what ways the various information is stored in the photographs is one of the most urgent questions that many researchers are working on nowadays.

The black hole shadow may be the primary characteristic of the image of a black hole. It has been argued that, from the precise shape of the black hole shadow, many physical parameters related to black holes can be read, such as the mass, spin and so on \cite{Takahashi:2005hy,Hioki:2009na,Tsupko:2017rdo,Kumar:2018ple,Tamburini:2019vrf,Bambi:2019tjh,Dokuchaev:2019pcx,Gralla:2020yvo,Gralla:2020srx}. Moreover, it was suggested in some previous works  that wormholes and other ultra-compact objects (UCOs) could have similar but different shadow curves from a black hole, which may provide a new observational way to distinguish a black hole from a wormhole or other UCOs \cite{Shaikh:2018lcc,Dai:2019mse,Joshi:2020tlq,Wang:2020emr,Paul:2020ufc,Dey:2020bgo,Wielgus:2020uqz,Zhang:2020xub}. In addition, some fundamental problems can be also probed with the help of the study of black hole shadows as some literature pointed out. These problems include but are not limited to the extreme environment of black holes, dark matters,  accretion disk, accelerating universe, modified gravity and extra dimensions, see examples in \cite{Bambi:2008jg,Johannsen:2010ru,Amarilla:2011fx,Grenzebach:2015oea,Amir:2016cen,Abdujabbarov:2016hnw,Dastan:2016vhb,Wang:2017hjl,Cunha:2018acu,Wang:2018eui,
Hennigar:2018hza,Perlick:2018iye,Bisnovatyi-Kogan:2018vxl,Vagnozzi:2019apd,Banerjee:2019nnj,Tsupko:2019mfo,Ovgun:2018tua,Wang:2018prk,Wei:2019pjf,Kumar:2019ohr,
Shaikh:2019fpu,Contreras:2019nih,Ovgun:2019jdo,Contreras:2019cmf,Wang:2019skw,Das:2019sty,Zhang:2019glo,Lu:2019zxb,Dokuchaev:2019jqq,Feng:2019zzn,
Kumar:2019pjp,Ma:2019ybz,Kumar:2020hgm,Li:2020drn,Chang:2020miq,Wei:2020ght,Pantig:2020uhp,Roy:2020dyy,Guo:2020zmf,Chen:2020aix,
EslamPanah:2020hoj,Zeng:2020dco,Ovgun:2020gjz,Jusufi:2020cpn,Badia:2020pnh,Belhaj:2020nqy,Khodadi:2020jij, Allahyari:2019jqz, Khodadi:2020gns, Ghosh:2020tdu, Chang:2020lmg,Belhaj:2020rdb,
Zeng:2020vsj,Dokuchaev:2020wqk,Peng:2020wun,Belhaj:2020mlv,Belhaj:2020okh,Psaltis:2020lvx,Contreras:2020kgy,Gralla:2020pra,Cotaescu:2020kcr,
Volkel:2020xlc,Guo:2018kis,Yan:2019etp,Guo:2019pte,Guo:2019lur,Li:2020val,Guo:2020qwk}. These interesting questions have greatly stimulated the study of black hole shadows, both theoretically and experimentally.

Theoretically,  the black hole shadows depend largely on our understanding of photon trajectories in the black hole spacetime. In a vacuum, photons move along geodesics under the approximation of geometric optics, thus the equations of motion of photons are completely governed by the background of the spacetime. However, it is always inevitable that there is some other fields besides the gravitational field in a real spacetime. Thus, a possibly realistic scenario is that photons, as a kind of electromagnetic field, interact with other fields as they travel outside the black hole. In particular, the magnetic fields play a crucial role in the black hole physics \cite{Abramowicz:2011xu} and there found some evidences showing that magnetic fields may exist around supermassive black holes. For example, in the vicinity of SgrA* and M87 the intensity of the magnetic field could be as large as 100 G \cite{Eatough:2013nva,Doeleman:2012zc}. Therefore, the magnetic field should be taken into account in order to get closer to the real physical situation. The path of the photon will deviate from the geodesic in the presence of a magnetic field when considering the quantum electrodynamics (QED) effects of photons.  It was shown that including the QED effect gives rise to the birefringence phenomenon \cite{DeLorenci:2000yh,Novello:1999pg}, leading to the deviation from the geodesics.
Consequently, the change of the trajectory of the photons certainly will change the shape of the black hole shadow observed by an observer\footnote{As for the interaction between photons and other fields, such as axion-like particles, some interesting works have been made and some exciting results were found in  \cite{McDonald:2019wou,Schwarz:2020jjh,Chen:2020qyp}.}.

In this paper, we would like to investigate the QED effect on black hole shadow, due to the background magnetic fields. We will consider three different magnetic field configurations. We first start with the static black hole with magnetic monopoles. We find an analytical expression when the effective coupling between the photon and magnetic field is small.  Next we study carefully the shadow of a Schwarzschild black hole in a magnetic field by employing the numerical backward ray-tracing method. Moreover,  we extend the study to the case that the black hole is surrounded by a current loop which generates magnetic field.  We find some remarkable features that may help to determine the mass or magnetic field outside the black hole in the future EHT experiments.

The plan of our paper is organized as follows: in section \ref{section2}, we review the dispersion relations induced by the QED effect. In section 3, we study the static black holes with magnetic monopoles. In section 4, we investigate the shadow of a Schwarzschild black hole in a uniform magnetic field at first and discuss the extension to a Schwarzschild black hole with a current loop. In section 5, we summarize our results. In addition, we  introduce the numerical backward ray-tracing method  carefully in appendices.

In this work, we have set the fundamental constants $c$, $G$ and the vacuum permittivity $\varepsilon_0$ to unity, and we will work in the convention $(-, +, +, +)$.

\section{Dispersion relations}\label{section2}

In QED, the one-loop vacuum polarization gives rise to the Euler-Heisenberg
effective Lagrangian \cite{Heisenberg:1935qt} for the electromagnetic field. For low frequencies, i.e.
$\omega \ll m_e$, this is \cite{Drummond:1979pp}
\begin{equation}
  L_{eff} = - \frac{1}{4} F_{\mu \nu} F^{\mu \nu} - \frac{\mu}{2} \left[ \frac{5}{4} (F_{\mu \nu} F^{\mu \nu})^2 - \frac{7}{2} F_{\mu
  \nu} F_{\sigma \tau} F^{\mu \sigma} F^{\nu \tau} \right],
\end{equation}
where $\mu=\frac{\hbar e^4}{360\pi^2  m_e^4}$ with $m_e$ being the electron mass  and $e$ being the charge of the electron \cite{Ahmadiniaz:2020lbg, Karbstein:2016hlj, Karbstein:2019oej}. Note that in the geometric unit, $\hbar$ is not necessarily to be set to unity, instead one finds $\hbar=m_p^2$ where $m_p$ denotes the Planck mass.
Let us consider the case that electromagnetic gauge potential is minimally coupled to gravity, which
leads to the following action
\begin{equation}\label{action}
  I = \int \sqrt{- g} d^4 x \left( \frac{1}{16 \pi} R + L_{eff}
  \right).
\end{equation}
The Einstein's equation is given by
\begin{equation}
  R_{\mu \nu} - \frac{1}{2} R g_{\mu \nu} = 8 \pi  T_{\mu \nu},
\end{equation}
where the energy momentum is
\begin{equation}
  T_{\mu \nu} = g_{\mu \nu} L_{eff} + F_{\mu \rho} F_{\nu}^{^{\,\,
  \rho}} - \frac{\mu}{2} \left[ - 5 F_{\alpha \beta}^{} F^{\alpha
  \beta} F_{\mu \rho} F_{\nu}^{^{\,\,\rho}} + 14 F_{\mu}^{\,\, \sigma}
  F_{\sigma \rho} F_{}^{\rho \tau} F_{\tau \nu} \right],
\end{equation}
and the modified \ Maxwell equation becomes
\begin{equation}
  \nabla_{\mu} F^{\mu \nu} + \mu [5 \nabla_{\mu}
  (F_{\alpha \beta}^{} F^{\alpha \beta} F^{\mu \nu}) - 14 \nabla_{\mu} (F^{\mu
  \sigma} F_{\sigma \tau} F^{\nu \tau})] = 0.
\end{equation}
Consider a beam of electromagnetic wave is propagating around the magnetic
black hole, which is equivalent to introducing a small perturbation
$\hat{A}_{\mu}$ into this background. Namely, there is $A_\mu=A^0_\mu+\hat{A}_{\mu}$, where $A^0_\mu$ induces the background magnetic field. Usually, when the wavelength $\lambda$ of the
electromagnetic wave  is much smaller than the characteristic
scale of the background, such as the radius $r_h$ of the event horizon of the
black hole, i.e. $\lambda / r_h \ll 1$, which indeed occurs for
celestial bodies including black holes and neutral stars, then the so-called
geometric optics approximation is valid. Under the geometric optics approximation,  the light lays is described
by the following Hamiltonian
\begin{equation}
  H = (p^2 - 8 \mu E^2) (p^2 - 14 \mu E^2),
\end{equation}
where
\begin{equation}
  E^{\mu} = F^{\mu \nu} p_{\nu}.
\end{equation}
From the Hamiltonian one obtains two branches of dispersion relation \cite{DeLorenci:2000yh},
\begin{equation}
  p^{}_{\alpha} p_{\beta} g^{\alpha \beta} + \lambda F^{\mu \alpha} p_{\alpha}
  F_{\mu}^{\,\, \beta} p_{\beta} = 0,
\end{equation}
with
\begin{equation}
  \lambda = - 8 \mu, \mathrm{or} - 14 \mu .
\end{equation}
From a different perspective, the dispersion relations can be interpreted as
that of the null vector $p_{\mu}$ with the geometry described by the following
effective metric \cite{Novello:1999pg}
\be\label{effectivemetric}
  G_{\alpha \beta}= g_{\alpha \beta} + X_{\alpha\beta},
\ee
where we have defined a new tensor $X_{\alpha\beta}\equiv\lambda F^{\mu}_{\,\, \alpha} F_{\mu \beta}$.

\section{Black hole with magnetic monopole}

In this section, let us start from the simplest case: a static black hole with magnetic monopole.  In this case, we ignore  the backreaction of the nonlinear QED correction to the geometry, and the energy momentum tensor is from pure Maxwell action. Now  the stationary Kerr-Newmann  is
described by the  metric
\begin{equation}
  d s^2 = - \frac{\Delta}{\Sigma} (d t - a \sin^2 \theta d \phi)^2 +
  \frac{\sin^2 \theta}{\Sigma} ((r^2 + a^2) d \phi - a d t)^2 +
  \frac{\Sigma}{\Delta} d r^2 + \Sigma d \theta^2,
\end{equation}
where
\begin{equation}
  \Delta = r^2 - 2 M r + (Q_e^2 + Q_m^2) + a^2, \quad \Sigma = r^2 + a^2
  \cos^2 \theta .
\end{equation}
$M$ and $a$ denotes the mass and angular momentum per unit mass, $Q_e$ and $Q_m$
are the electric and magnetic charges. The associated one-form potential has
nonvanishing components
\begin{equation}
  A_{\mu} d x^{\mu} = \frac{Q_e r - Q_m a \cos \theta}{\Sigma} d t + \frac{-
  Q_e a r \sin^2 \theta + Q_m (r^2 + a^2) \cos \theta}{\Sigma} d \phi .
\end{equation}

For simplicity,  we  focus on the spherically symmetric black hole, and leave the case of Kerr in a forthcoming paper \cite{Zhong2020}. When the black hole is static and only magnetic charges are considered, the metric can be written as
\begin{equation}\label{RNqm}
  d s^2 = - f (r) d t^2 + \frac{d r^2}{f (r)} + r^2 (d \theta^2 + \sin^2
  \theta d \phi^2), \quad f (r) = 1 - \frac{2 M}{r} + \frac{Q_m^2}{r^2},
\end{equation}
and the gauge field is
\begin{equation}
  A_{\mu} d x^{\mu} = Q_m \cos \theta d \phi .
\end{equation}
In this case, one finds that the effective metric (\ref{effectivemetric}) is simply given by
\begin{equation}
  \ds = - f (r) d t^2 + \frac{d r^2}{f (r)} + h (r) (d \theta^2 + \sin^2
  \theta d \phi^2), \quad h = r^2 + \frac{\lambda Q_m^2}{r^2} .
\end{equation}
Since the black hole is \ static and spherically symmetric, one can always
restrict the photons in the equatorial plane, i.e. $\theta = \pi / 2$. The
dispersion relation can be explicitly written as
\begin{equation}
  - f (r) \dot{t}^2 + \frac{\dot{r}^2}{f (r)} + h \dot{\phi}^2 = 0.
\end{equation}
Moreover, let $p^\mu$ denote the 4-momentum of the photons, we use $q_\mu=G_{\mu\nu}p^\nu$ to represent the dual vector of $p^\mu$ in terms of the effective spacetime. Then the two conserved quantities, energy and angular momentum takes the form of
\begin{equation}
  E = - q_t, \quad L = q_{\phi},
\end{equation}
which allows us to write the orbit equation as
\begin{equation}
  \left( \frac{d r}{d \phi} \right)^2 = V_{eff},
\end{equation}
with the effective potential
\begin{equation}
  V_{eff} = h^2 \left( \frac{E^2}{L^2} - \frac{f}{h} \right) .
\end{equation}
By evaluating the equations,
\begin{equation}
  V_{eff} = 0, \quad V_{eff}' = 0,
\end{equation}
we can obtain the radius of the circular null geodesic, which is determined by
\begin{equation}\label{photons}
  \frac{f' (r_{ph})}{f (r_{ph})} - \frac{h' (r_{ph})}{h
  (r_{ph})} = 0.
\end{equation}
In general the solution of this equation is very complicated. However, to study the effects of QED qualitatively, we assume the the coupling constant $\mu$ is small. Here and hereafter, we would like to set $M=1$ for simplicity. As we work in the unit system $c=G=\varepsilon_0=1$, $\mu$ has the dimension of mass square and the magnetic field has the dimension of inverse mass. Their combination $\mu B^2$ is dimensionless, which characterizes the strength of the QED effect. In the case of black hole with magnetic monopole, by setting $M=1$ we can find that the smaller black hole, the larger $\mu$. In other words, the smaller black hole yields the larger $\mu B^2$ around the horizon, such that the nonlinear correction of the QED effect gets enhanced.
In this case, the variation of $\mu$ can reflect the change of the mass of the black hole.
In contrast, for a black hole embedded in a uniform magnetic field, as we will discuss in next section, the QED effect overall is independent of the black hole mass.

For a supermassive black hole, the magnetic field around it is small such that one can neglect  the nonlinear QED effect safely. Up to the leading order in
$\mu$, one obtains
\begin{equation}\label{appph}
  r_{ph} = R_{ph} + \delta\lambda,
\end{equation}
where
\be
R_{ph}=\frac{1}{2}\left(3+\sqrt{9-8Q_m^2}\right)
\ee
is the photon sphere radius for the RN black hole with no QED effect and
\be
\delta=\frac{Q_m^2\left(R_{ph}-1\right)}{8Q_m^4+27R_{ph}-18Q_m^2\left(R_{ph}+1\right)}
\ee
describes the deviation from the result without considering the QED effects. Due to the spherical symmetry, the photons will fill all the circular orbits
to form a so-called photon sphere. The corresponding constant of motion $b=L /
E$ for this photon sphere is given by
\begin{equation}\label{appbc}
  b_c =B_{c}+\Delta\lambda.
\end{equation}
where
\bea
B_{c}&=&\frac{R_{ph}^2}{\sqrt{R_{ph}-Q_m^2}}\nn\\
\Delta&=&\frac{Q_m^2}{\sqrt{2\left(R_{ph}-Q_m^2\right)}\sqrt{8Q_m^4+54R_{ph}-12Q_m^2\left(2R_{ph}+3\right)}}
\eea
are the critical impact parameter and the deviation from the result  for the RN black hole without QED effect. The existence of unstable photon sphere means the appearance of the observable of the black hole, the black hole shadow. Consider all null geodesics that go from the position of the static observer at $(t_o, r_o, \pi / 2, \phi_o= 0)$ into the past. The critical null geodesics that orbit around the black hole on the photon sphere will leave the observer at an angle $\theta$ with respect to
the radial line that satisfies
\begin{equation}
 \tan \theta =\left.\frac{r d \phi}{g_{r r} d r}\right\vert_{r_o} .
\end{equation}
This angle describes the angular size of the shadow of the black hole. For a static observer at large distance, i.e., $r_o\gg 1$, the radius of the black hole shadow is given by
\begin{equation}
 r_{sh}=r_o\sin\theta=b_c.
\end{equation}
Noticing that $\lambda<0$ for both branches of dispersion relation, we conclude that the radius of the black hole shadow becomes smaller when we include the QED effect. And the branch corresponding to $\lambda=-14\mu$ has smaller shadow size compared to the one with $\lambda=-8\mu$.

\begin{figure}[t!]
\begin{center}
\includegraphics[width=80mm,angle=0]{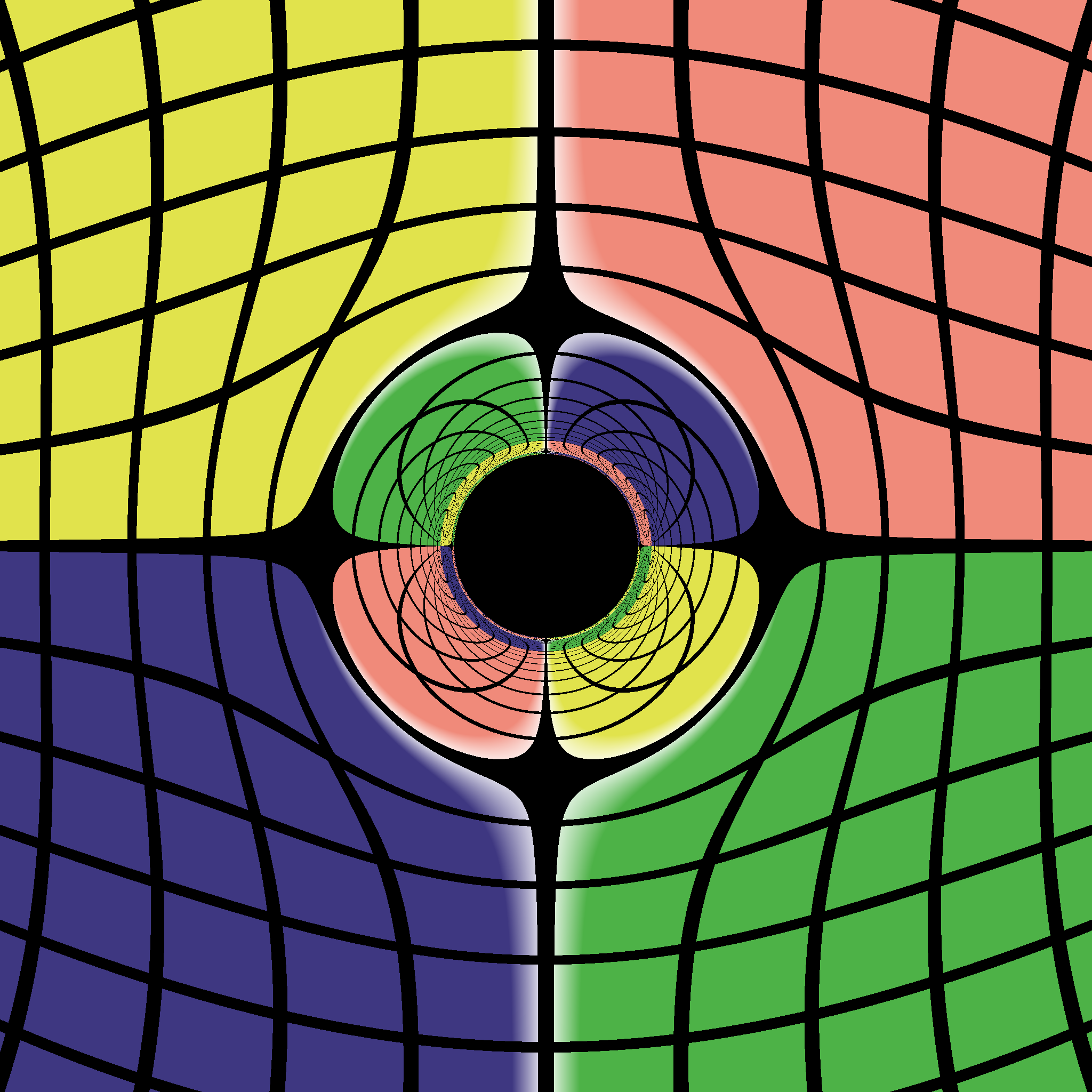}
\end{center}
\caption{The shadow of the static and magnetically charged black hole including the QED effect using the numerical ray-tracing method. We set the black hole mass $M=1$, the magnetic charge $Q_m=0.1$ and the coupling constant $\mu=1$ with $\lambda=-8\mu$. }\label{sn815}
\end{figure}

In addition to the approximate expressions in Eq. (\ref{appph}) and (\ref{appbc}), it is  possible to study the case that the mass of the black hole is not so large that the effective coupling constant $\mu\sim 1$ and the effect of QED becomes significant.\footnote{We work in the window that the nonlinear QED effect does not affect the background geometry but is important for the photon trajectory.} However, since the black hole is spherically symmetric and we can always choose $\theta=\pi/2$, the variables of the geodesic equations for photons can be separated. Thus, the radius of the black hole shadow can be calculated straightforwardly even though Eq. (\ref{photons}) is complicated. 

\begin{figure}[t!]
\begin{center}
\includegraphics[width=160mm,angle=0]{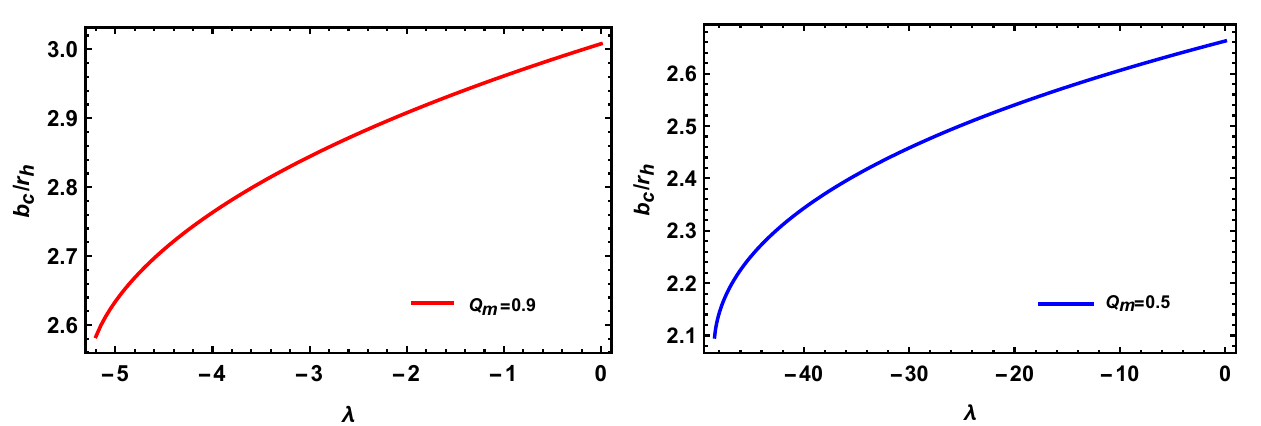}
\end{center}
\caption{The variation of the shadow radius with respect to $\lambda$ for the static black hole with magnetic monopole including the QED effect. In the graph of the left, we set the magnetic charge $Q_m=0.9$, and $Q_m=0.5$ for the right side. The range of effective coupling $\lambda$ is determined by the  constraint $h>0$ outside the black hole. The values of $\lambda$ is related to the actual masses of the black holes. }\label{qmlam}
\end{figure}

Instead, we give an example of the black hole shadow in Fig. \ref{sn815} using numerical backward ray-tracing method. Moreover, we present the variation of the black hole shadow radius with respect to $\lambda$ in Fig. \ref{qmlam}. The causality of the effective metric requires $\partial_t$ is always timelike and the other translational vectors along the coordinates are spacelike outside the black hole, that is, $h>0$ has to be hold for $r>2$, which gives us a constraint on $\lambda$. From Fig. \ref{qmlam}, we find $\lambda$ has a wider range of values when $Q_m$ becomes smaller. On the other hand, we find with the increase of $|\lambda|$, the ratio of the radius of black hole shadow to the event horizon radius becomes smaller. In addition, it should be pointed out that one may need consider the backreaction of the nonlinear QED effect when $|\lambda|$ in Fig. \ref{qmlam} is very large, even though the range of $\lambda$ we consider here is consistent with $h>0$ under our initial assumptions. Recall that the effective coupling $\lambda$ is inversely proportional to the mass square of the black hole, the above fact suggest that for a smaller black hole  the shadow radius is nearer to the black hole horizon. This phenomenon is most obvious for extreme black hole.  

\begin{figure}[h!]
  \centering

  \subfigure[$\Lambda=0.00$]{
  \begin{minipage}[t]{0.3\linewidth}
  \centering
  \includegraphics[width=1.5in]{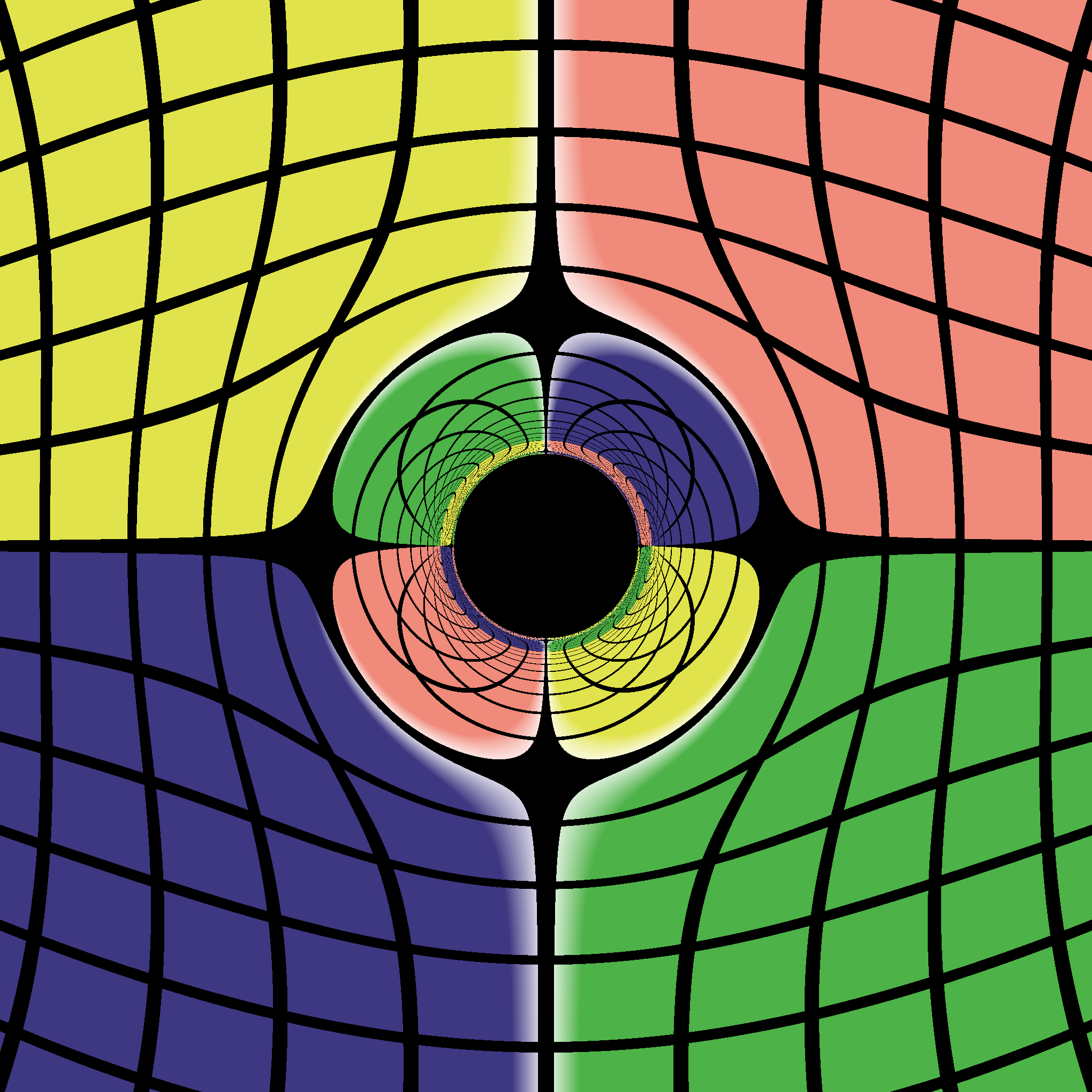}
  \end{minipage}%
  }%
  \subfigure[$\Lambda=-0.20$]{
  \begin{minipage}[t]{0.3\linewidth}
  \centering
  \includegraphics[width=1.5in]{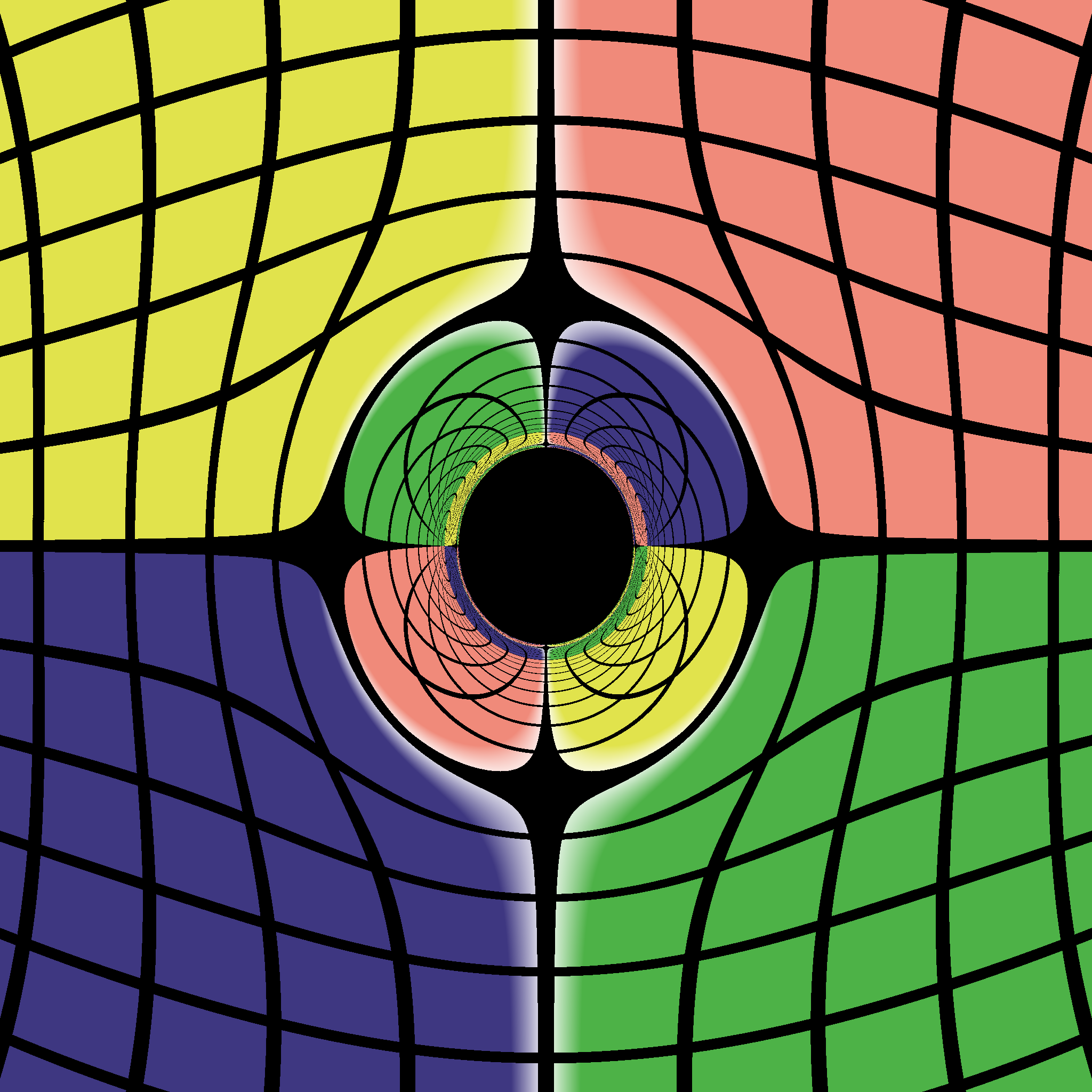}
  \end{minipage}%
  }%
  \subfigure[$\Lambda=-0.40$]{
  \begin{minipage}[t]{0.3\linewidth}
  \centering
  \includegraphics[width=1.5in]{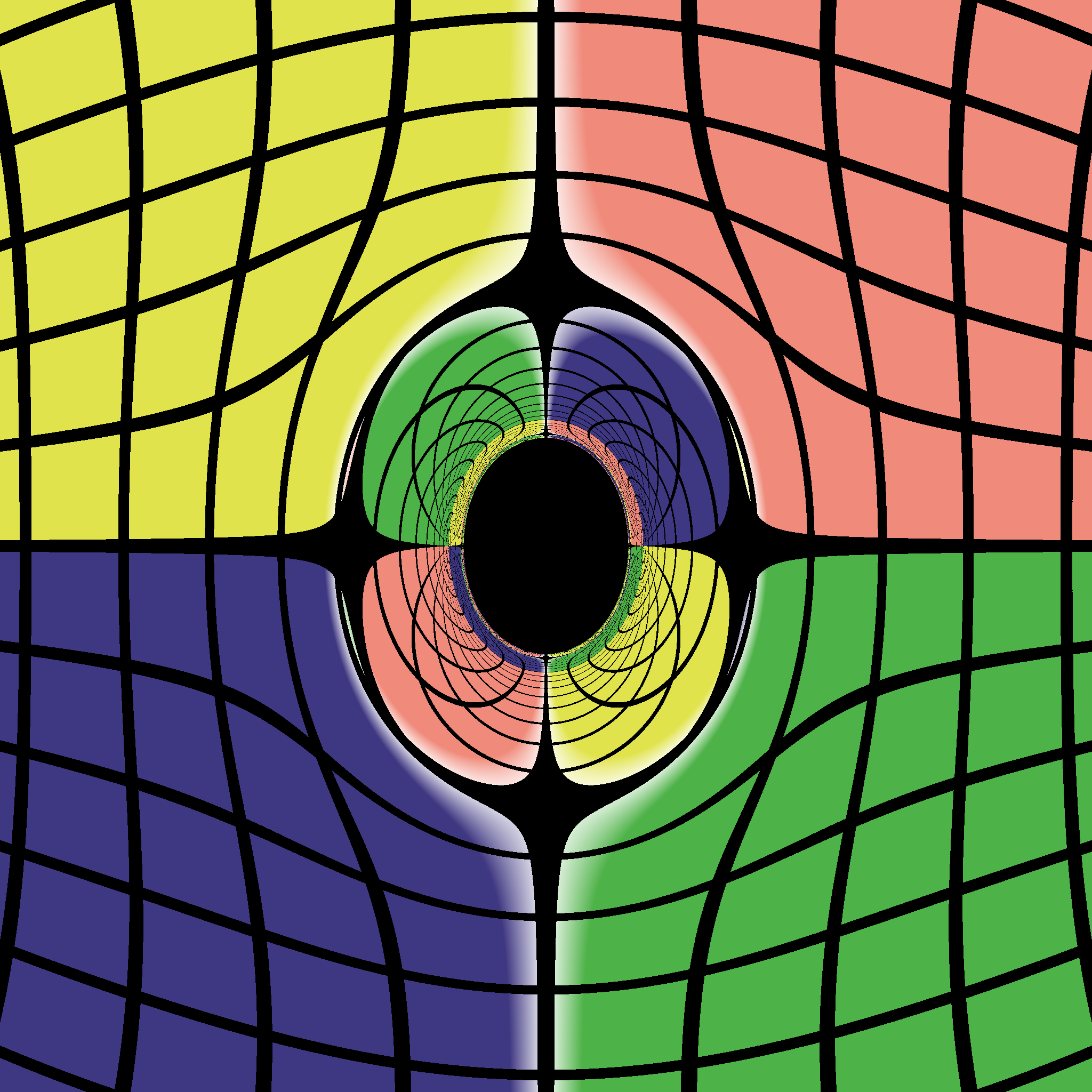}
  \end{minipage}
  }%

  \subfigure[$\Lambda=-0.60$]{
  \begin{minipage}[t]{0.3\linewidth}
  \centering
  \includegraphics[width=1.5in]{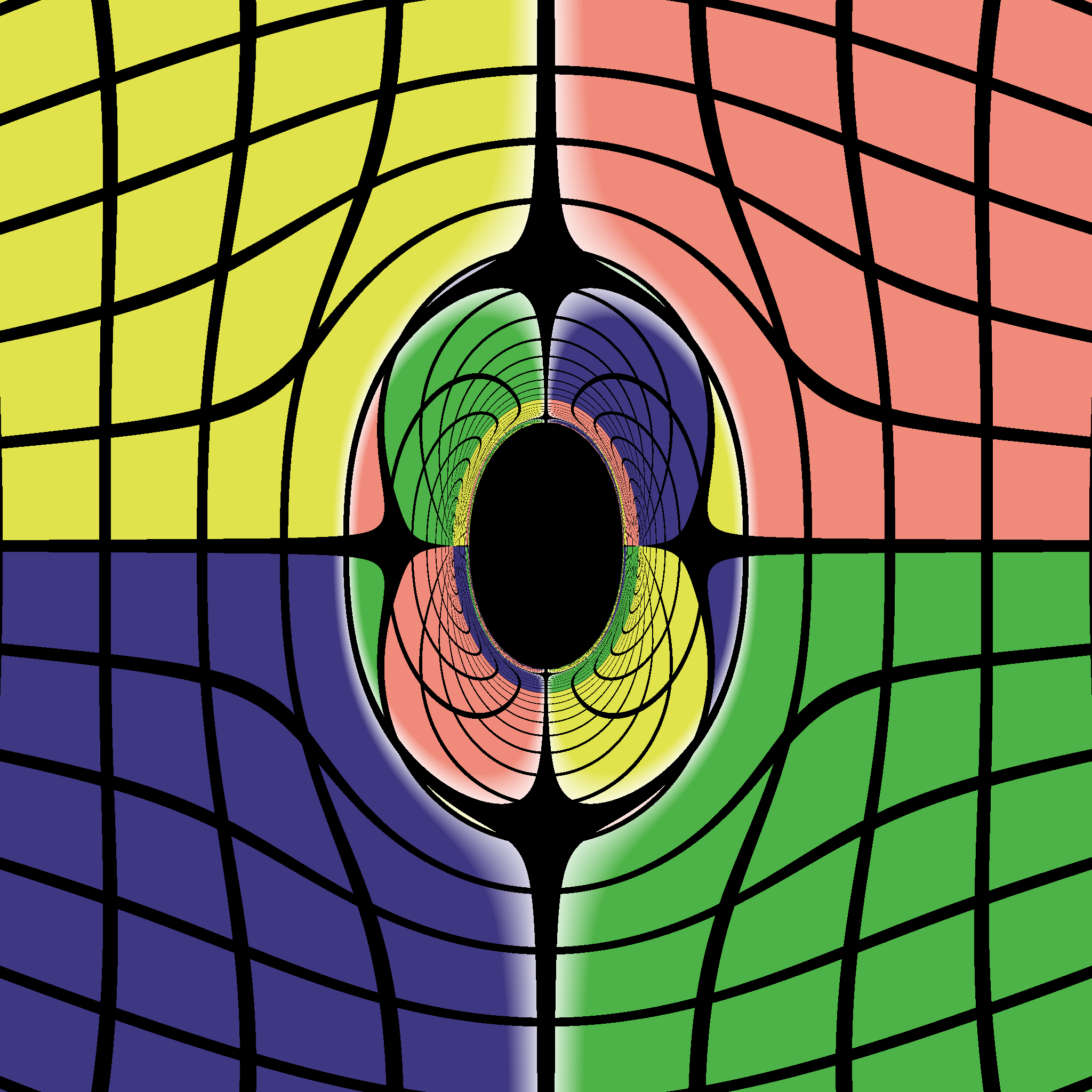}
  \end{minipage}
  }%
  \subfigure[$\Lambda=-0.80$]{
  \begin{minipage}[t]{0.3\linewidth}
  \centering
  \includegraphics[width=1.5in]{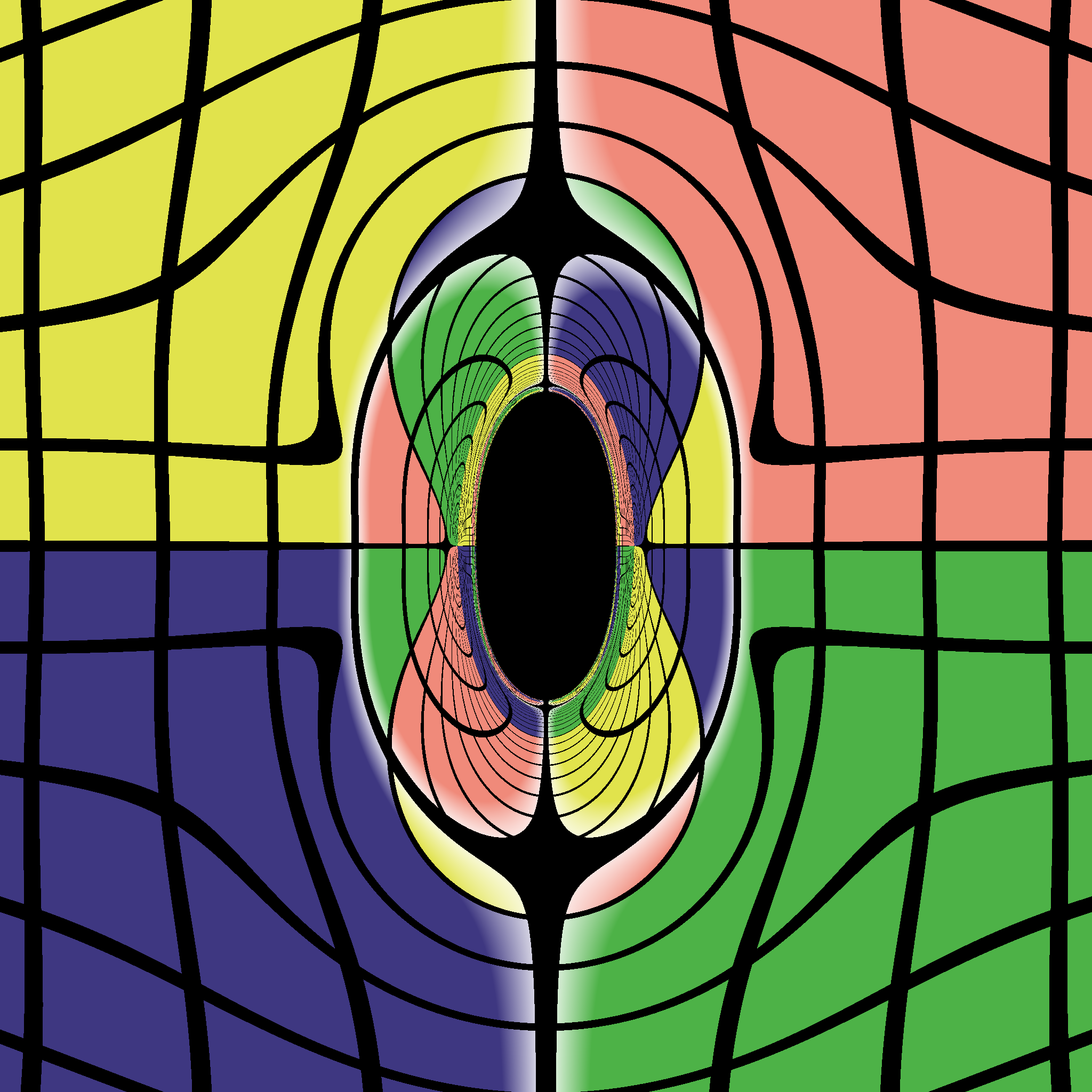}
  \end{minipage}
  }%
  \subfigure[$\Lambda=-0.95$]{
  \begin{minipage}[t]{0.3\linewidth}
  \centering
  \includegraphics[width=1.5in]{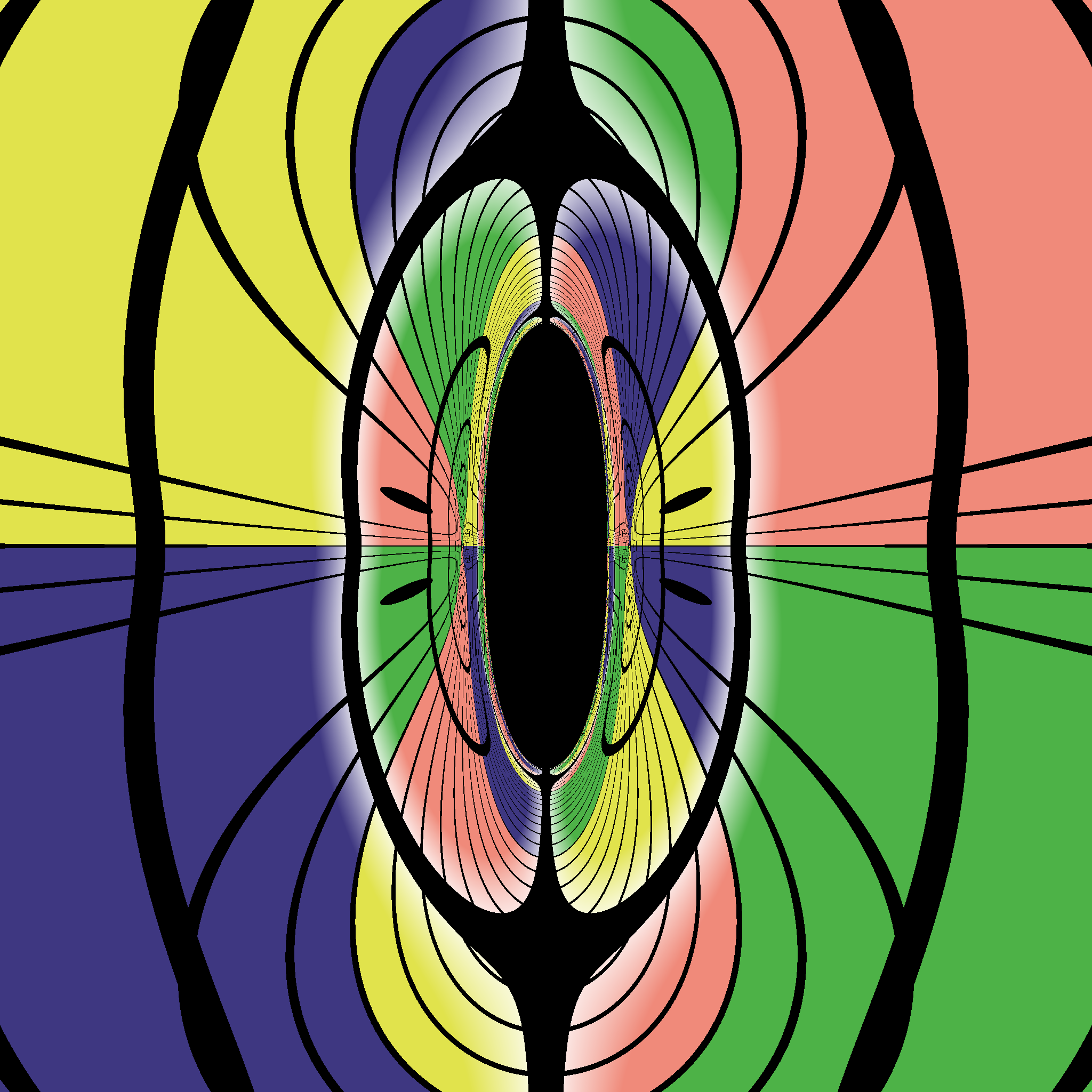}
  \end{minipage}
  }%

  \centering
  \caption{The images of the  static  black hole in uniform magnetic fields. The inclination angle of the observer is fixed at $\theta_o=\pi/2$.}
  \label{uniform1}
\end{figure}

\section{Neutral black holes in magnetic fields}

In this section, we turn to consider a more realistic case that a neutral black hole is bathed in magnetic fields, at least in the vicinity of the black hole. Before we get into the specific magnetic field models, let us  have some general discussions. As we know, the magnetic vector is related to electromagnetic field tensor by
\begin{equation}\label{magneticfield}
  B^i = \varepsilon^{t i j k} F_{j k}, \quad i = r, \theta, \phi,
\end{equation}
where the repeated index does not mean summation. Here, we consider the case where there is no electric field, thus we have $A_t=0$. Since $A_\mu$ is a gauge field, we can always take the gauge choice $A_r=0$. Hence, the general form of the gauge field reads
\be
A_\mu dx^\mu=A_\theta(r,\theta) d\theta+A_\phi(r,\theta)d\phi.
\ee
where we have assumed the gauge field is independent of $\phi$  for a stationary spacetime of our interest. 

\begin{figure}[h!]
  \centering

  \subfigure[$\theta_o=\pi/20$]{
  \begin{minipage}[t]{0.3\linewidth}
  \centering
  \includegraphics[width=1.5in]{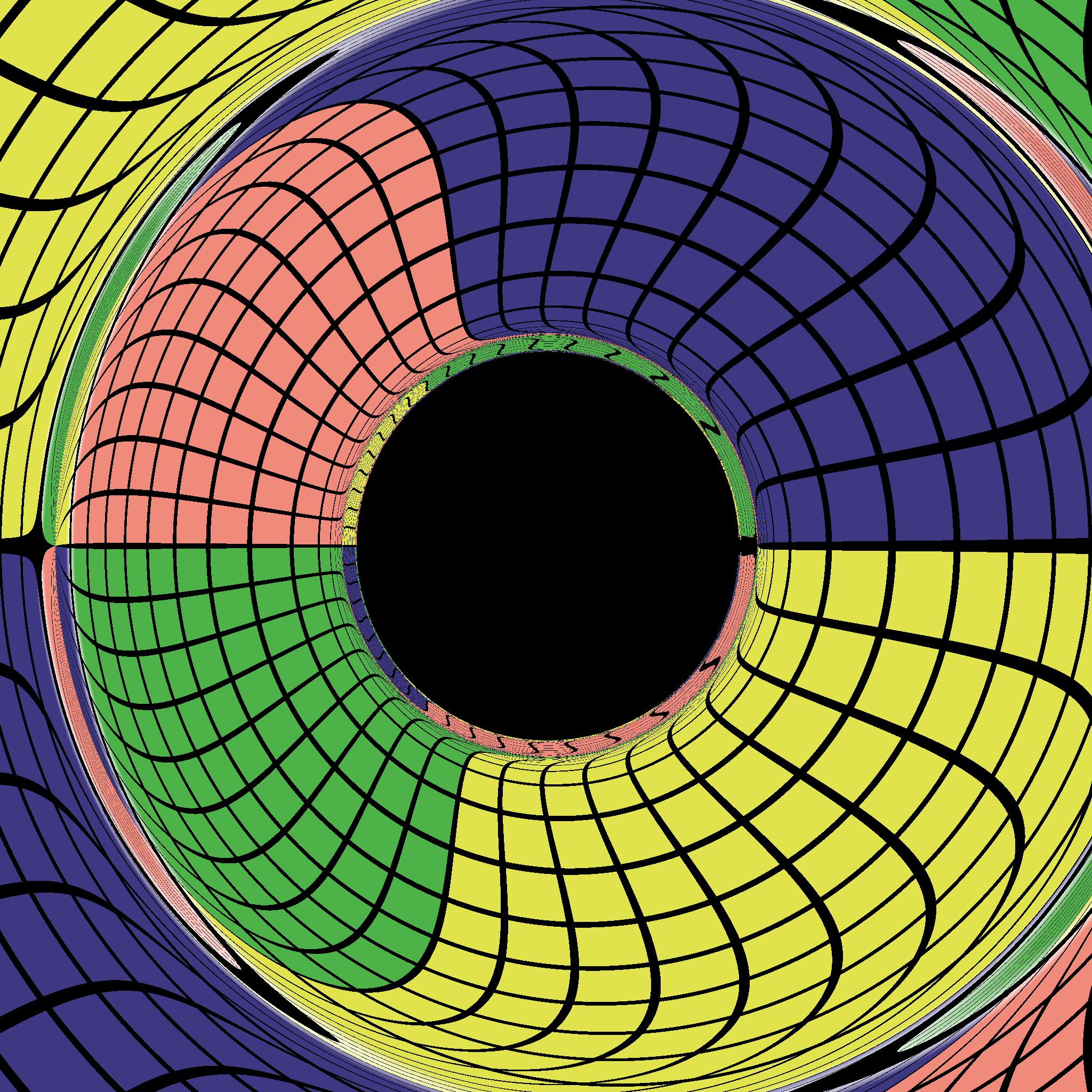}
  \end{minipage}%
  }%
  \subfigure[$\theta_o=\pi/4$]{
  \begin{minipage}[t]{0.3\linewidth}
  \centering
  \includegraphics[width=1.5in]{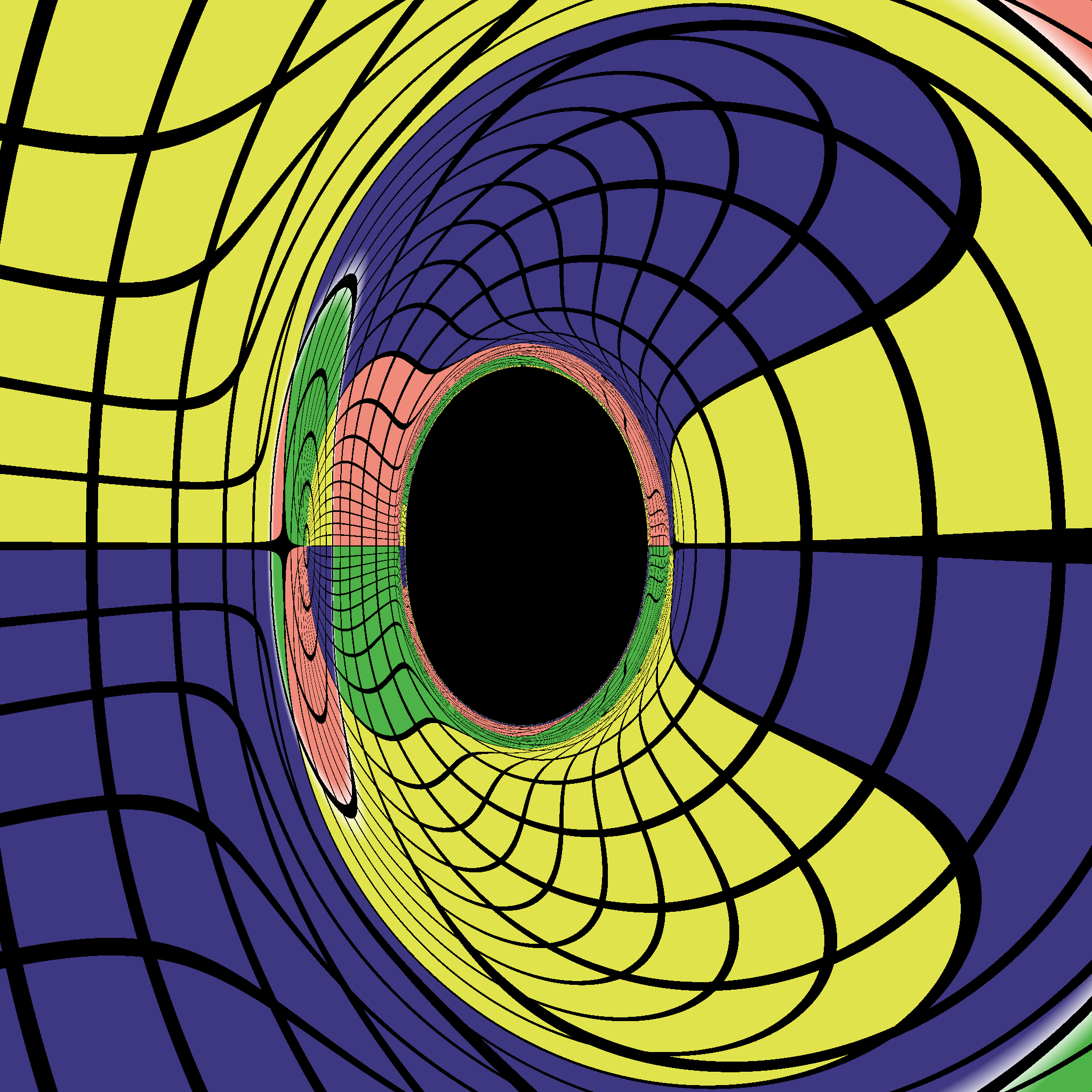}
  \end{minipage}%
  }%

  \subfigure[$\theta_o=\pi/2$]{
  \begin{minipage}[t]{0.3\linewidth}
  \centering
  \includegraphics[width=1.5in]{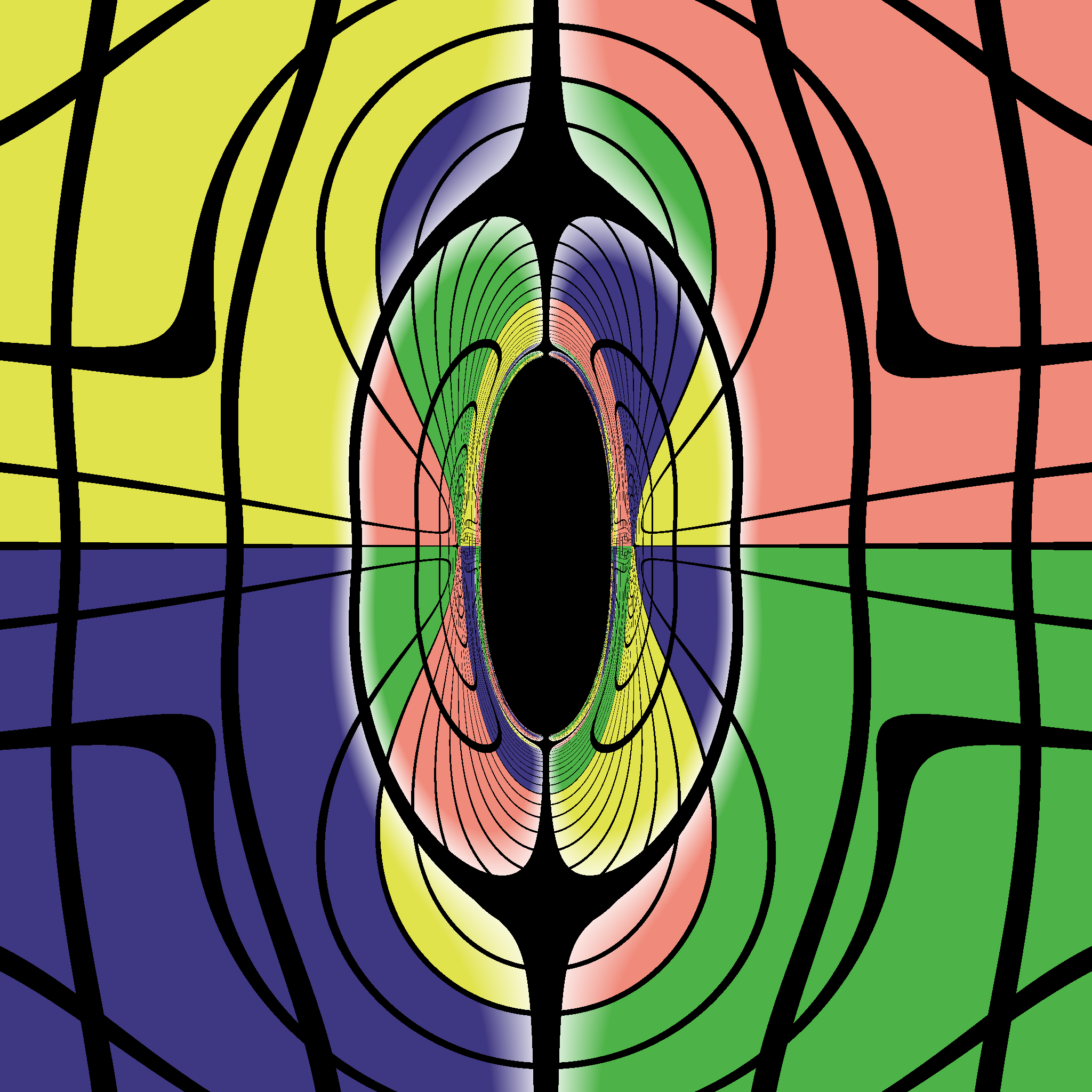}
  \end{minipage}
  }%
  \subfigure[$\theta_o=3\pi/4$]{
  \begin{minipage}[t]{0.3\linewidth}
  \centering
  \includegraphics[width=1.5in]{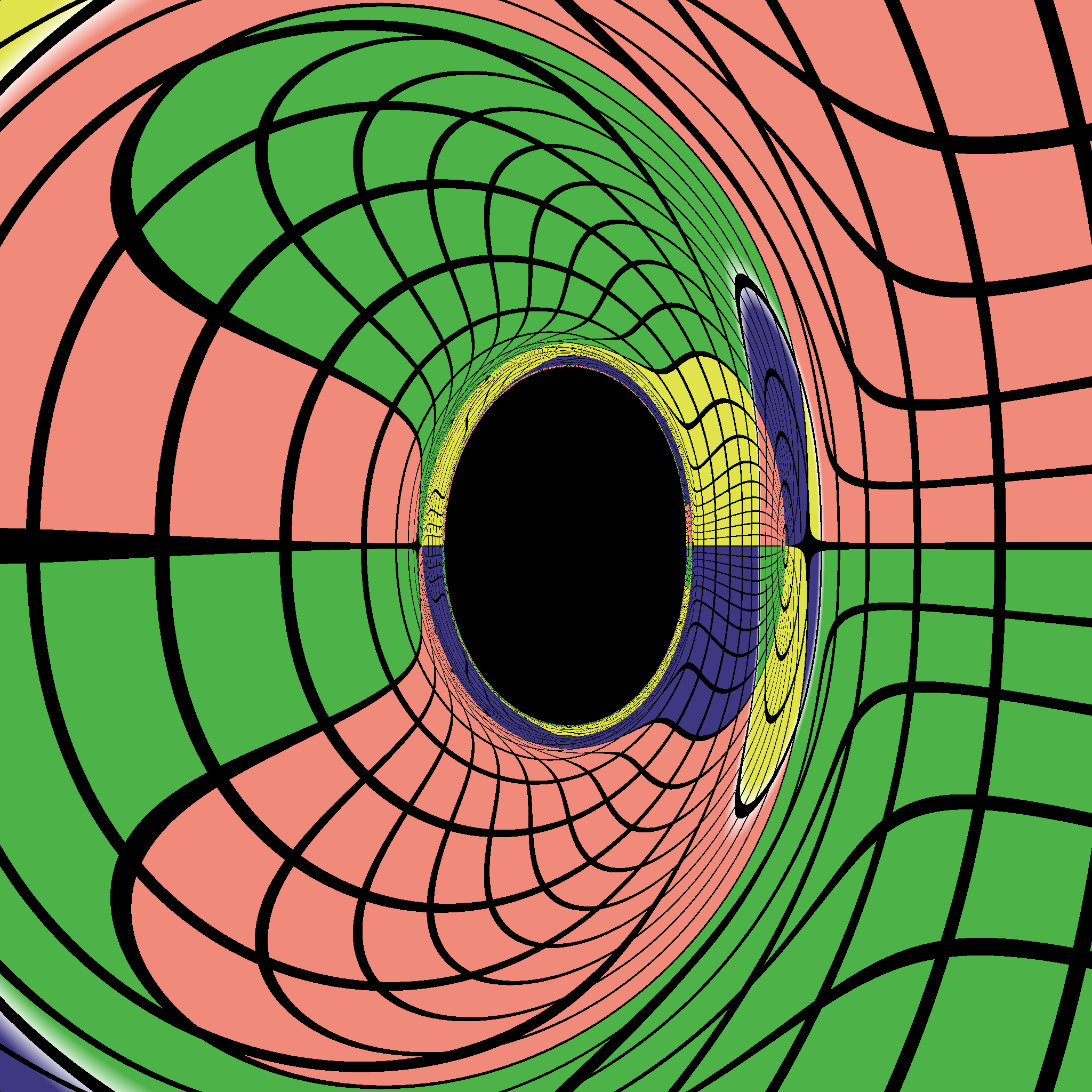}
  \end{minipage}
  }%
  \subfigure[$\theta_o=19\pi/20$]{
  \begin{minipage}[t]{0.3\linewidth}
  \centering
  \includegraphics[width=1.5in]{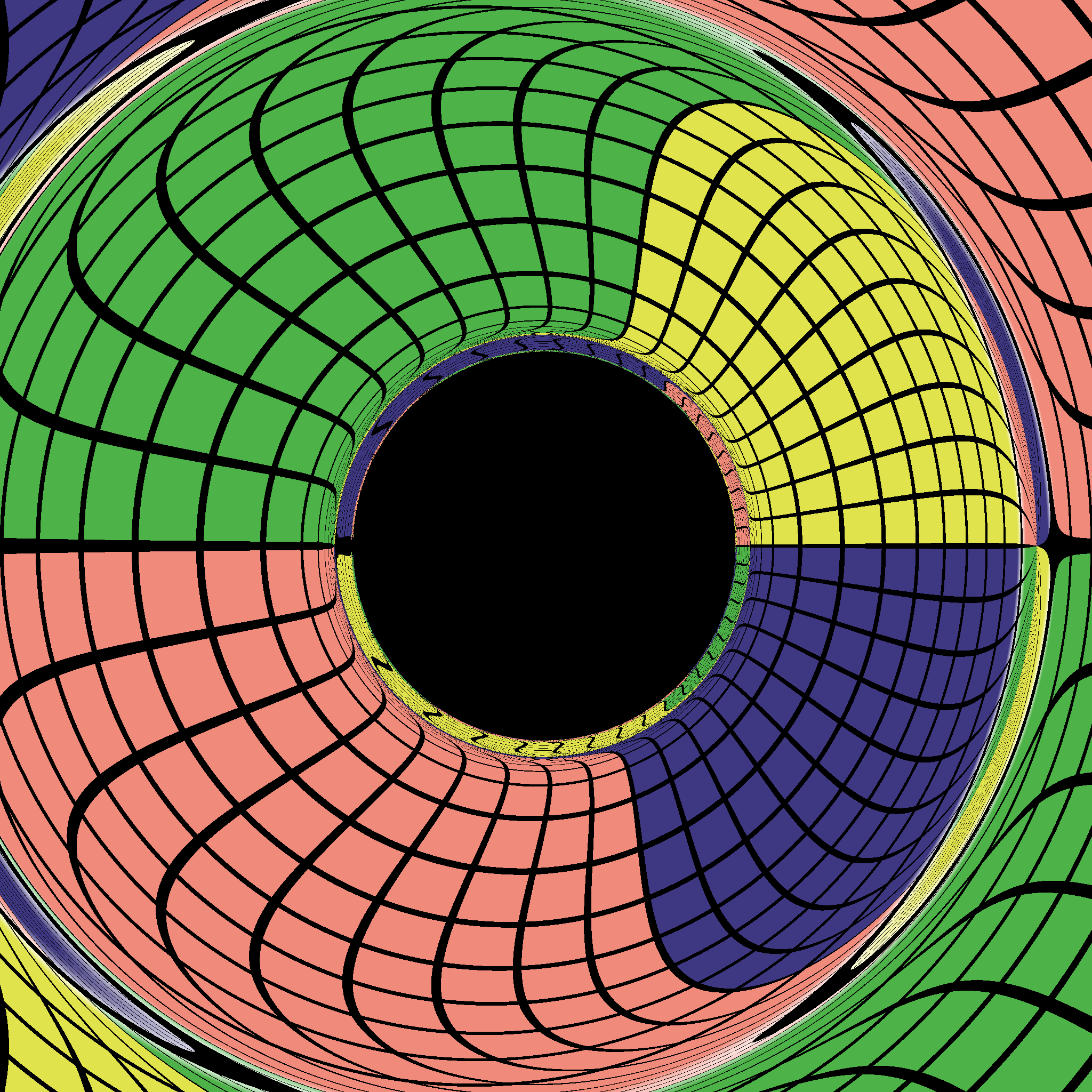}
  \end{minipage}
  }%

  \centering
  \caption{The images of the  static black hole in a uniform magnetic field with various inclination angles of the observer. $\Lambda$ is fixed at $-0.7$.}
  \label{uniformtheta}
\end{figure}

\subsection{Schwarzschild black hole in the uniform magnetic field}
In this subsection, we start from a simple case, that is, the gauge filed $A_\mu$ satisfy the source-free Maxwell equations
\be
\nabla_\mu F^{\mu\nu}=0.
\ee
It can be seen that this equation is solved by a gauge potential proportional to the Killing vector in a Ricci flat spacetime. Namely, a  Killing vector in a vacuum spacetime would generate a solution of source-free Maxwell's equations.  Thus, for a stationary spacetime of interest, the only non-zero component is $A_\phi$. In fact, this case has been studied by Wald in 1974 \cite{Wald:1974np}, and the solution to the source-free Maxwell equation in a static spacetime was found to be

\begin{equation}
  A_{\mu} d x^{\mu} = \frac{B}{2} r^2 \sin^2 \theta d \phi,
\end{equation}
where $B$ is the strength of the uniform magnetic field. Due to the presence of
magnetic field, the whole spacetime is not spherically symmetric, instead, it
is axisymmetric. One can expect that in this case, the shadow as seen by a
distant observer is no longer circular anymore, however, this spacetime is non-spinning, which implies the shadow may not be similar to that of a Kerr black hole, either. To investigate the shadow of this spacetime is very interesting.  


 \begin{figure}[t!]
 \begin{center}
 \includegraphics[width=150mm,angle=0]{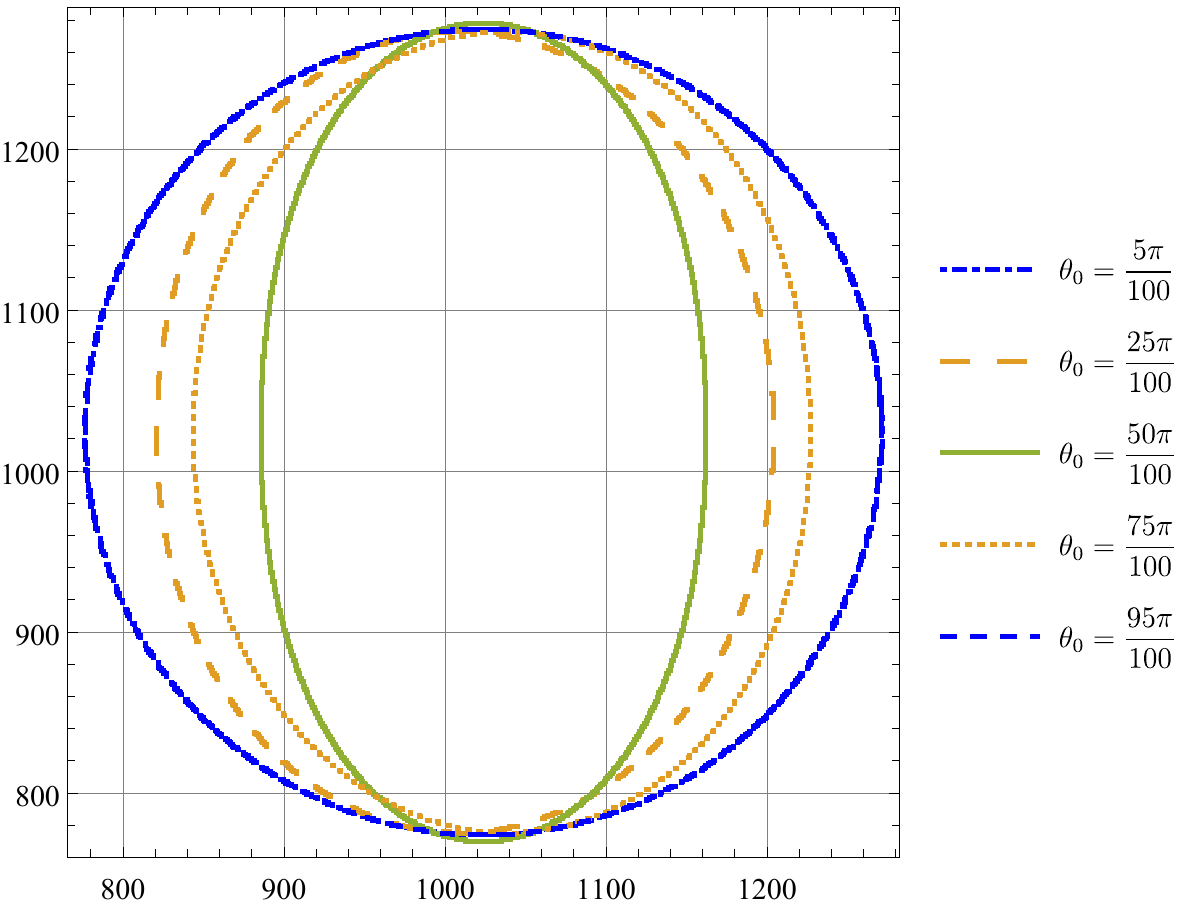}
 \end{center}
 \caption{The outline of the shadow of the  static black hole in a uniform magnetic fields with various inclination angles of the observer. Here $\Lambda=-0.7$ which is same with that in Fig. \ref{uniformtheta}.}\label{uniform9theta}
 \end{figure}

From (\ref{effectivemetric}) we find that the effective metric for the photon motion is given by
\begin{eqnarray}
  d s_{\text{eff}}^2 & = & - f (r) d t^2 + \left( \frac{1}{f(r)} + \lambda B^2 \sin^2 \theta
  \right) d r^2 + 2 \lambda B^2 r \sin \theta \cos \theta d r d \theta \\
  &  & + r^2 (1 + \lambda B^2 \cos^2 \theta) d \theta^2 + r^2 \sin^2 \theta
  (1 + \lambda B^2 (f (r) \sin^2 \theta + \cos^2 \theta)) d \phi^2, \nonumber
\end{eqnarray}
where $f(r)=1-\frac{2}{r}$ and $r=2$ is still the coordinate singularity, that is, the horizon is unchanged for the effective metric. In addition, notice that $\lambda$ always appears with $B^2$, we define a new parameter $\Lambda\equiv\lambda B^2$ for simplicity. The metric components $G_{rr}$, $G_{\theta\theta}$, $G_{\phi\phi}$ should be positive outside the horizon, we conclude $0\le|\Lambda|<1$. Moreover,  the effective metric obviously has two Killing vector, $\partial_t$ and
$\partial_{\phi}$ such that  we can define two conserved quantities, the energy and angular
momentum $L$, as usual
\begin{equation}
  E = - q_t, \quad L = q_{\phi} .
\end{equation}
Plugging this into the dispersion relation, we examine the separability of the
Hamiltonian,
\begin{eqnarray}
  - \frac{E^2}{f} + g^{r r}_{} q_r^2 + 2 g^{r \theta}_{} q_{\theta}^{} q_r +
  g_{\theta \theta} q_{\theta}^2 + g_{}^{\phi \phi} L^2=0.
\end{eqnarray}
Due to
 the crossing term $g^{r\theta}q_r q_{\theta}$ in this equation, the Hamiltonian cannot be decoupled. In addition, we want to emphasize that this term cannot be avoided by coordinates transformation, since the effective metric is noncircular because of the presence of the toroidal magnetic field, which is  in the direction of the rotational Killing vector. In other words, $A_\phi\neq0$ must lead to the non-zero crossing term $g_{r\theta}$, which was proved by Carter\cite{Carter1972}. Hence, we haves to resort to the ``numerical backward ray-tracing'' method to investigate the shadow of such a black hole inevitably. Here, to make the article concise, we would like to show our results immediately and leave the method and techniques to the appendix.

In Fig. \ref{uniform1}, we show the images of the static and magnetically charged black hole for various $\Lambda$ when the observer is located on the equatorial plane. As expected, the shadow is a circle when $\Lambda=0$. When the QED effects get involved, the shadow is no longer a perfect circle and the deformation becomes large with the increase of $|\Lambda|$. Moreover, it's interesting to investigate the variations of null trajectories and shadow with the viewing angle since $g_{r\theta}$ exists in the effective metric. In Fig. \ref{uniformtheta}, we show the shadows of the black holes at $\theta_o=\pi/20, \pi/4, \pi/2, 3\pi/4, 19\pi/20$ by fixing $\Lambda=-0.7$. Obviously, except $\theta_o=\pi/2$, The segment with the longest distance between two points in the shadow is not horizontal. To get a better idea of the relationship between the shadow and the observational angle, let us put these shadows on the same screen in Fig. \ref{uniform9theta}, where the numbers on the horizontal and vertical coordinates represent the positions of the pixels. From Fig. \ref{uniform9theta}, we find the profiles are symmetric about the horizontal and vertical axes for each $\theta_o$. In addition, the curve for $\theta_o$ and the one for $\pi-\theta_o$ coincide. This fact suggests that the uniform magnetic field does not break the $\mathcal{Z}_2$ symmetry of the spacetime.

\begin{figure}[h!]
  \centering

  \subfigure[$\Lambda=-0.10$]{
  \begin{minipage}[t]{0.3\linewidth}
  \centering
  \includegraphics[width=1.5in]{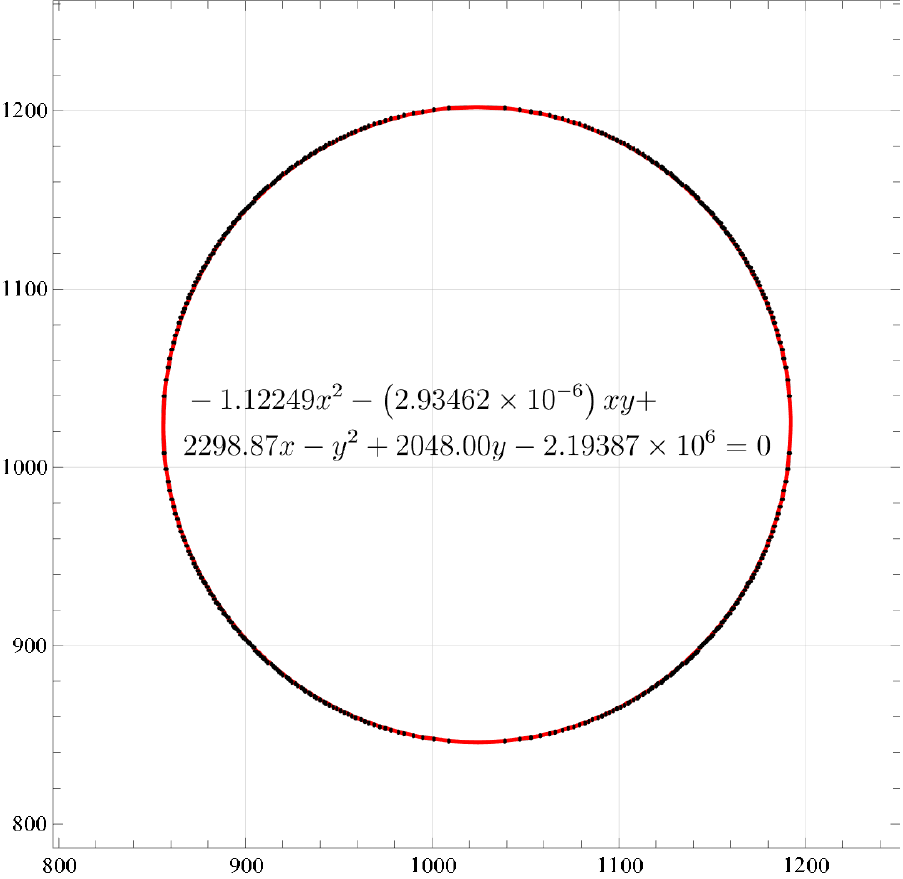}
  \end{minipage}%
  }%
  \subfigure[$\Lambda=-0.30$]{
  \begin{minipage}[t]{0.3\linewidth}
  \centering
  \includegraphics[width=1.5in]{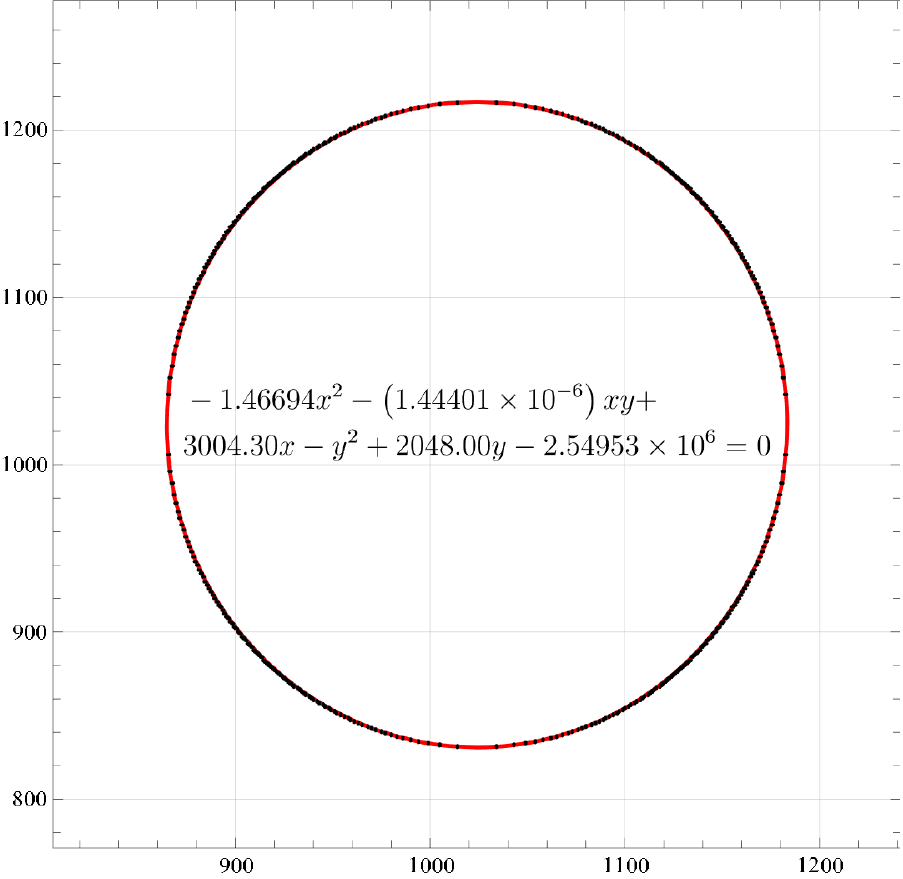}
  \end{minipage}%
  }%

  \subfigure[$\Lambda=-0.50$]{
  \begin{minipage}[t]{0.3\linewidth}
  \centering
  \includegraphics[width=1.5in]{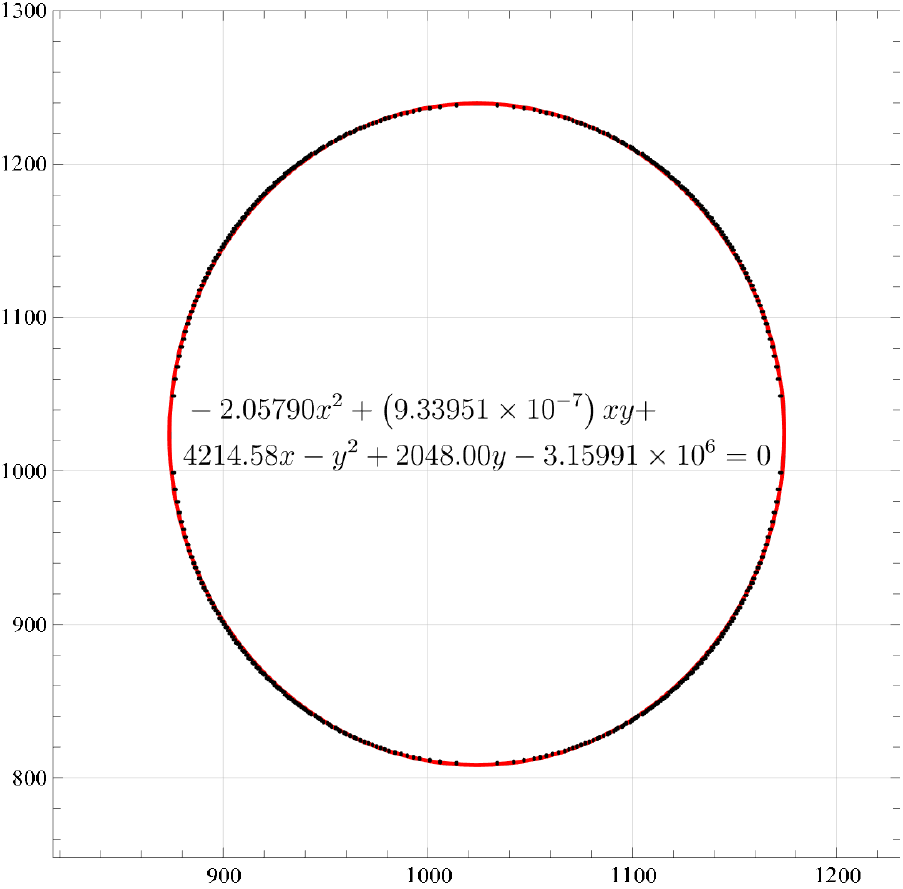}
  \end{minipage}
  }%
  \subfigure[$\Lambda=-0.70$]{
  \begin{minipage}[t]{0.3\linewidth}
  \centering
  \includegraphics[width=1.5in]{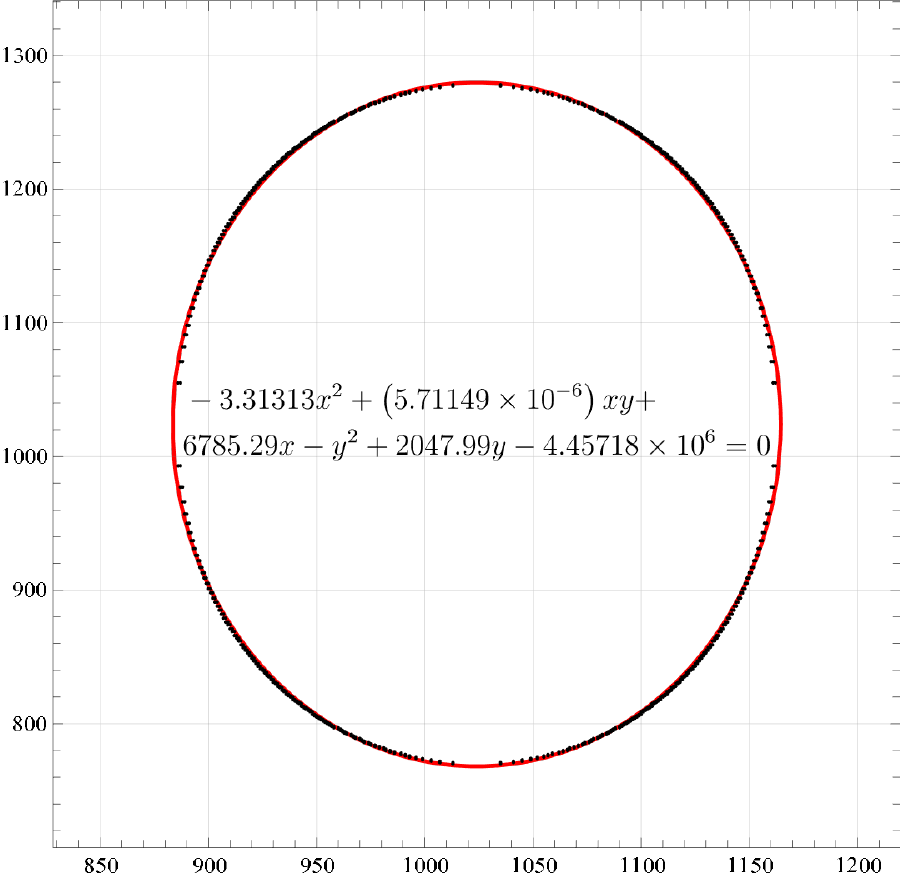}
  \end{minipage}
  }%
  \subfigure[$\Lambda=-0.90$]{
  \begin{minipage}[t]{0.3\linewidth}
  \centering
  \includegraphics[width=1.5in]{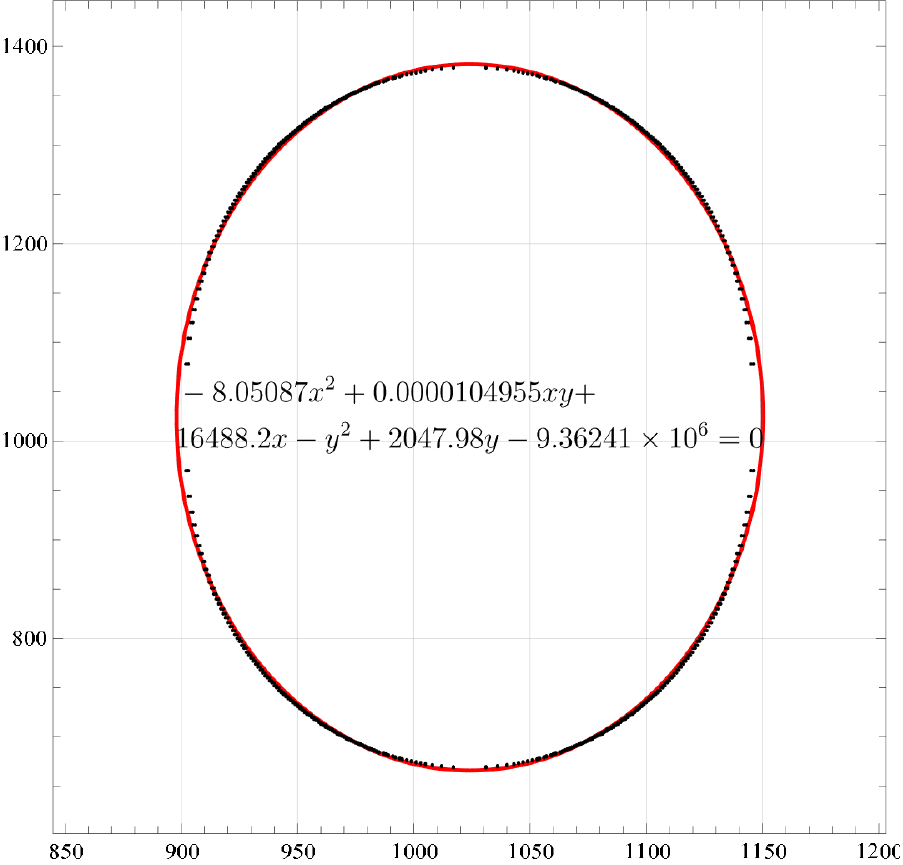}
  \end{minipage}
  }%

  \centering
  \caption{The images of the  static black hole in uniform magnetic fields. The inclination angle of the observer is fixed at $\theta_o=\pi/2$.}
  \label{uniformlambda}
\end{figure}

\begin{figure}[t!]
\begin{center}
\includegraphics[width=150mm,angle=0]{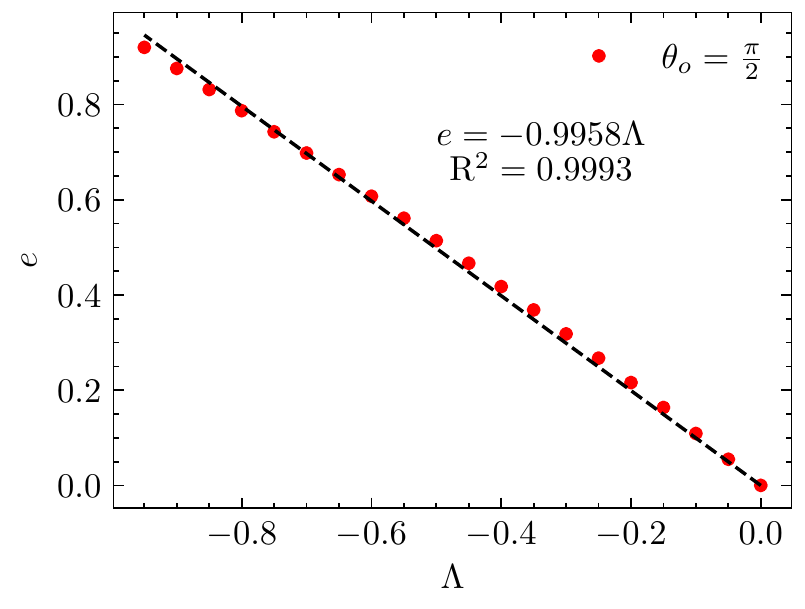}
\end{center}
\caption{The variation of the eccentricity with respect to $\Lambda$ for the static and magentically charged black hole. The inclination angle of the observer is fixed at $\theta_o=\pi/2$.}\label{uniforme}
\end{figure}
Another interesting question is whether the closed curve is an ellipse. To answer this question, let us first focus on the observer's position on the equatorial plane. In Fig. \ref{uniformlambda}, we show a series of shadow curves varying with $\Lambda$. The red curves are strict ellipses whose equations are presented in Fig. \ref{uniformlambda}. Using our numerical data, we draw the shadow curve with black dots. One can find that the black dots match the red ellipse very well when $|\Lambda|$ is not very large, say $|\Lambda|\le0.5$. When $\Lambda$ becomes large, only a few black points deviate from the ellipse a little bit. Thus, we think the shadow curves seen by an equatorial observer are  basically in oval shape. Moreover, we notice that the ellipse becomes flatter with a increasing $|\Lambda|$ qualitatively. In order to characterize this relationship more precisely, we introduce the eccentricity $e$ of an ellipse,  and in Fig, \ref{uniforme} we draw a series of points of $e$ and corresponding $\Lambda$. These data points fit nicely with a line, showing clearly a linear relation between the eccentricity and $\Lambda$. With this linear relation, we can infer the value of $\Lambda$ by measuring the shape of the black hole shadow using the EHT. Recall that $\Lambda=\lambda B^2$, we can further read the magnetic field strength around the  the black hole. 

\begin{figure}[h!]
  \centering

  \subfigure[$\theta_o=0.1\pi$]{
  \begin{minipage}[t]{0.3\linewidth}
  \centering
  \includegraphics[width=1.5in]{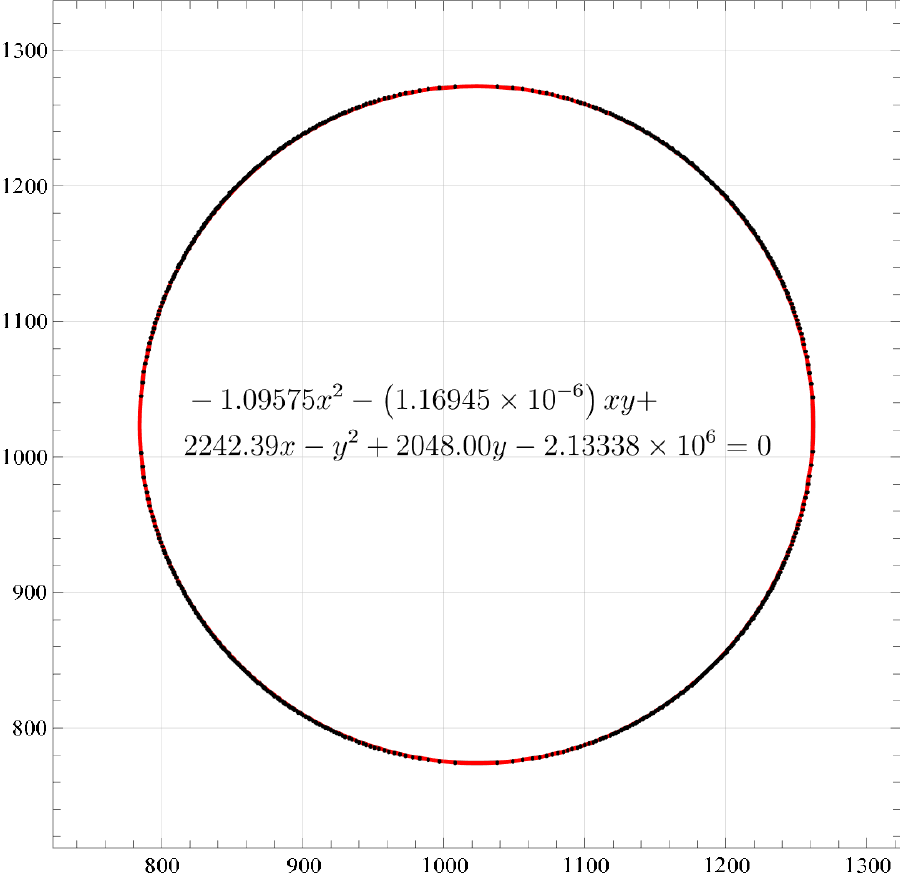}
  \end{minipage}%
  }%
  \subfigure[$\theta_o=0.3\pi$]{
  \begin{minipage}[t]{0.3\linewidth}
  \centering
  \includegraphics[width=1.5in]{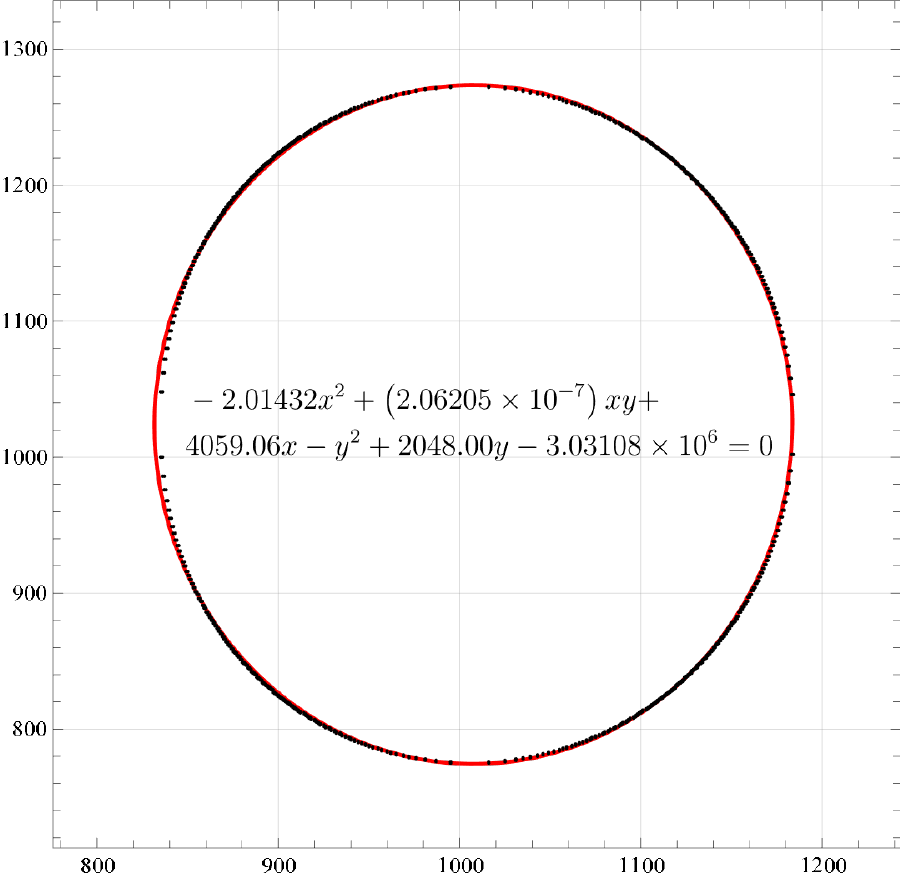}
  \end{minipage}%
  }%

  \subfigure[$\theta_o=0.5\pi$]{
  \begin{minipage}[t]{0.3\linewidth}
  \centering
  \includegraphics[width=1.5in]{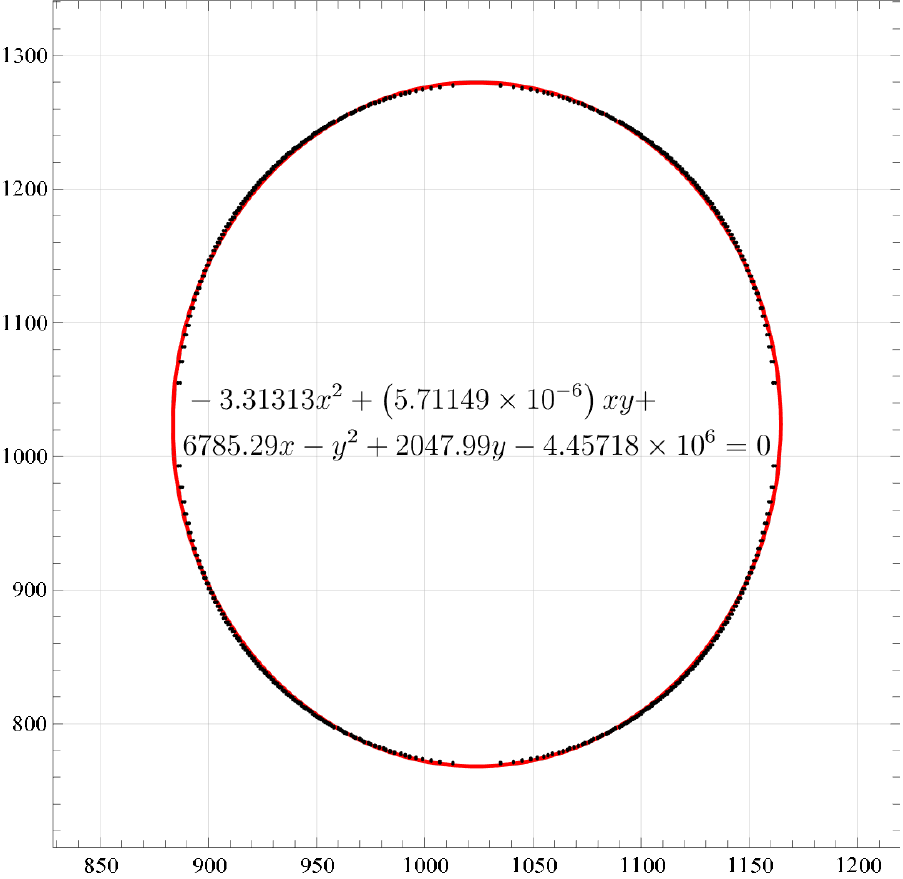}
  \end{minipage}
  }%
  \subfigure[$\theta_o=0.7\pi$]{
  \begin{minipage}[t]{0.3\linewidth}
  \centering
  \includegraphics[width=1.5in]{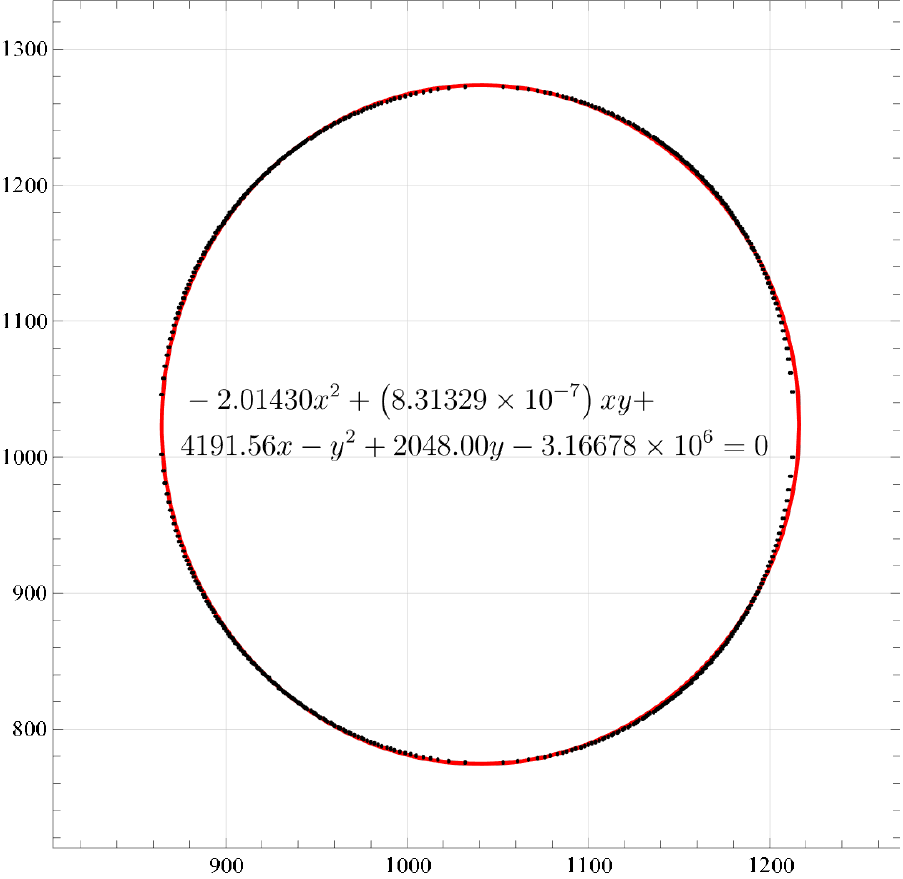}
  \end{minipage}
  }%
  \subfigure[$\theta_o=0.9\pi$]{
  \begin{minipage}[t]{0.3\linewidth}
  \centering
  \includegraphics[width=1.5in]{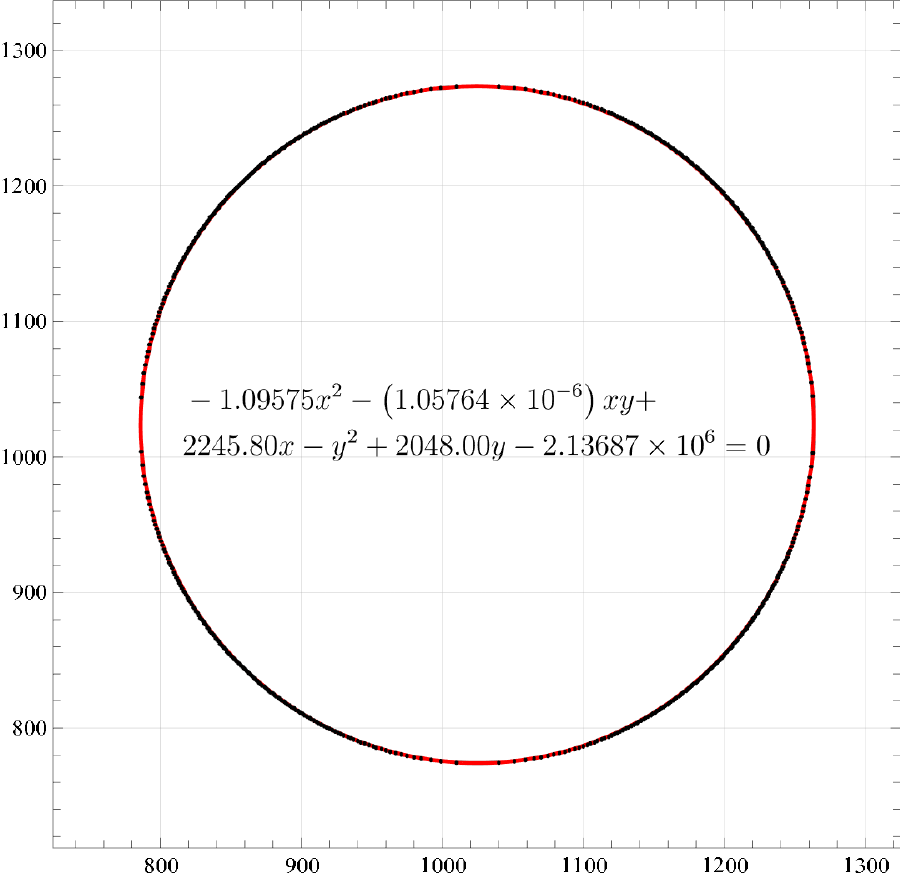}
  \end{minipage}
  }%

  \centering
  \caption{The shapes of the shadows of the  static black hole in uniform magnetic fields observed by  non-equatorial observers with $\Lambda=-0.7$.}
  \label{uniformthetao}
\end{figure}

\begin{figure}[t!]
\begin{center}
\includegraphics[width=150mm,angle=0]{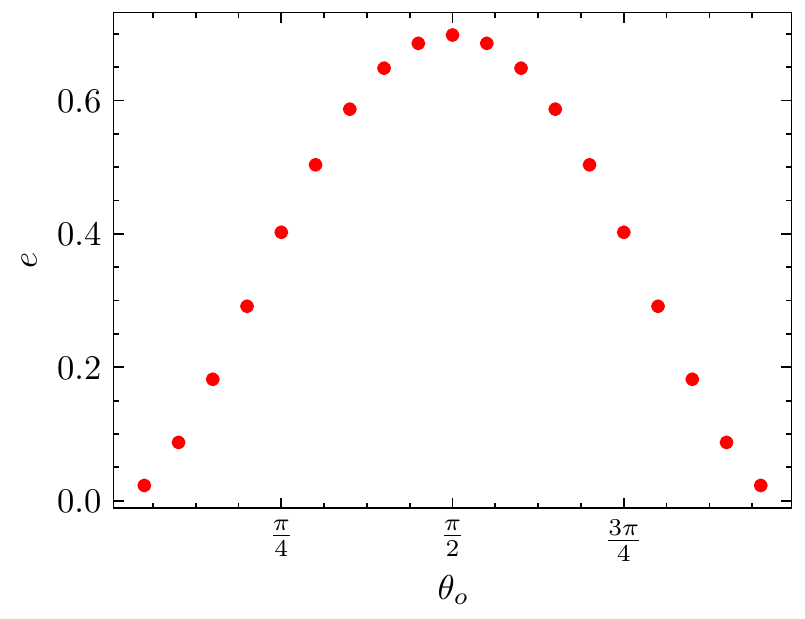}
\end{center}
\caption{The variation of the eccentricity with respect to various $\theta_o$ for the static black hole in a uniform magnetic field with $\Lambda=-0.7$.}\label{uniformt}
\end{figure}

Next, we turn to the shape of the shadow curve observed by non-equatorial observers. Similarly, by fitting the data to ellipses, we find only a few data deviate from ellipses when $\theta_o$ is near $0$, $\pi/2$, $\pi$, see the examples in Fig. \ref{uniformthetao}. This is reasonable since the shadow curve nearly circular when the observer faces up or faces down. For an observer with $\theta_o$ being near $\pi/4$ or $3\pi/4$, even though the shape of the shadow looks still like an    ellipsis  for $|\Lambda|=0.7$, the deviation from an ellipsis becomes significant  for $|\Lambda|=0.9$.  Thus the shadow curves can be approximately treated as ellipses in most cases.  We then show the variation of the eccentricity with respect to $\theta_o$ with $\Lambda=-0.7$ in Fig.\ref{uniformt}.

\subsection{A current loop around a Schwarzschild Black hole}
In this subsection, we turn to a relatively complicated situation in which there exists a source on the right hand of the Maxwell equation, that is,
\be
\nabla_\mu F^{\mu\nu}=4\pi J^\nu
\ee
Such kind of a solvable model was proposed by J. A. Petterson in 1974 \cite{Petterson:1974bt}. The current takes in this form
\begin{equation}
J^\phi = \frac{I}{R_c^2} \sqrt{1-\frac{2}{R_c}} \delta(r-R_c) \delta(cos \theta)
\end{equation}
where, $R_c$ is the radius of the loop, and $I$ is the current density set to be $1$ later in this paper. Note that the current has axisymmetry and there is no electric field involved, thus only $A_\phi(r, \theta)$ is non-vanishing. The solution of the Maxwell equation can be expanded in terms of spherical harmonic functions: for the gauge field in the radius $R_c$, it is
\begin{equation}\label{innerf}
A_\phi=\sum_{l=1}^\infty \alpha_l[r P_l(r-1)P_l^1(\cos \theta) \sin\theta - \frac{r^2-2r}{l(l+1)} P_{l,r} (r-1)P_l^1 (\cos \theta) \sin\theta ],
\end{equation}
while for the outer gauge field,
\begin{equation}\label{outerf}
A_\phi=\sum_{l=1}^\infty \beta_l[r Q_l(r-1)P_l^1(\cos \theta) \sin\theta - \frac{r^2-2r}{l(l+1)} Q_{l,r} (r-1)P_l^1 (\cos \theta) \sin\theta ]
\end{equation}
where $P_l$ and $Q_l$ are the first and second kind Legendre functions, and $P_{l,r}, Q_{l,r}$ are their derivatives with respect to $r$, $P_l^1$ is the associated Legendre function. It is worth noting that if all the terms are included, there appears a divergence when considering the QED effect. To fix this problem, a simple way is to keep finite orders of harmonics. And for the simplest case, we can drop all the terms except $l=1$, that is, only the dipolar term is kept. On the other hand, the continuity of $A_\phi$ should be respected at $r=R_c$. In the end,  we have
\begin{equation}
A_\phi=h(r) \sin^2\theta
\end{equation}
with
\begin{align}
&h(r)=
\begin{cases}\label{conti}
-3\pi \sqrt{1-\frac{2}{R_c}}V(R_c)r^2 & \text{  $2<r<R_c$}\\
-3\pi \sqrt{1-\frac{2}{R_c}}V(r)R_c^2 & \text{  $r>R_c$},\\
\end{cases} \\
&V(r)=\frac{1}{4}[r+1+\frac{r^2}{2} \log(1-\frac{2}{r})].
\end{align}\label{poten}
Here we ignore the backreaction of the magnetic field to the background. Moreover, we want to stress that, inside the current loop, the vector potential can be expressed in terms of a uniform magnetic field $B$ as
\be
A_\phi=\frac{1}{2}Br^2\sin^2\theta
\ee
with
\be
B=-6\pi \sqrt{1-\frac{2}{R_c}}V(R_c).
\ee
However, outside the current loop, the magnetic field $B$ is dependent on $r$, which takes in this form
\be
B_o(r)=-6\pi \sqrt{1-\frac{2}{R_c}}R_c^2\frac{V(r)}{r^2}.
\ee
It is convenient to introduce a new dimensionless quantity
\begin{align}
\kappa(r)=\frac{B(r)}{B}=
\begin{cases}
1 & \text{  $2<r<R_c$}\\
\frac{R_c^2}{V(R_c)}\frac{V(r)}{r^2} & \text{  $r>R_c$},\\
\end{cases}.
\end{align}
thus we have $A_\phi=\frac{1}{2}\kappa B r^2\sin^2\theta$ and the effective metric reads
\begin{eqnarray}
  d s_{eff}^2 & = & - f (r) d t^2 + \left( \frac{1}{f(r)} + \frac{\Lambda Z^2}{r^2\sin^2\theta}
  \right) d r^2 + \frac{2\Lambda ZW}{r^2\sin^2\theta}drd\theta \\
  &  & +  \left(r^2+ \frac{\Lambda W^2}{r^2\sin^2\theta}\right) d\theta^2 +\left( r^2 \sin^2 \theta
   +\Lambda f Z^2+\frac{\Lambda W^2}{r^2}\right) d \phi^2, \nonumber
\end{eqnarray}
where $\Lambda=\lambda B^2$, $Z=\kappa r\sin^2\theta$ and $W=\kappa r^2\sin\theta\cos\theta$. Note that the magnetic field $B$ is related to the position of the current loop  and this effective metric is asymptotically flat which are different from the effective metric discussed in the subsection 4.1.

\begin{figure}[h!]
  \centering

  \subfigure[$R_c=3, \Lambda=-0.33$]{
  \begin{minipage}[t]{0.5\linewidth}
  \centering
  \includegraphics[width=2in]{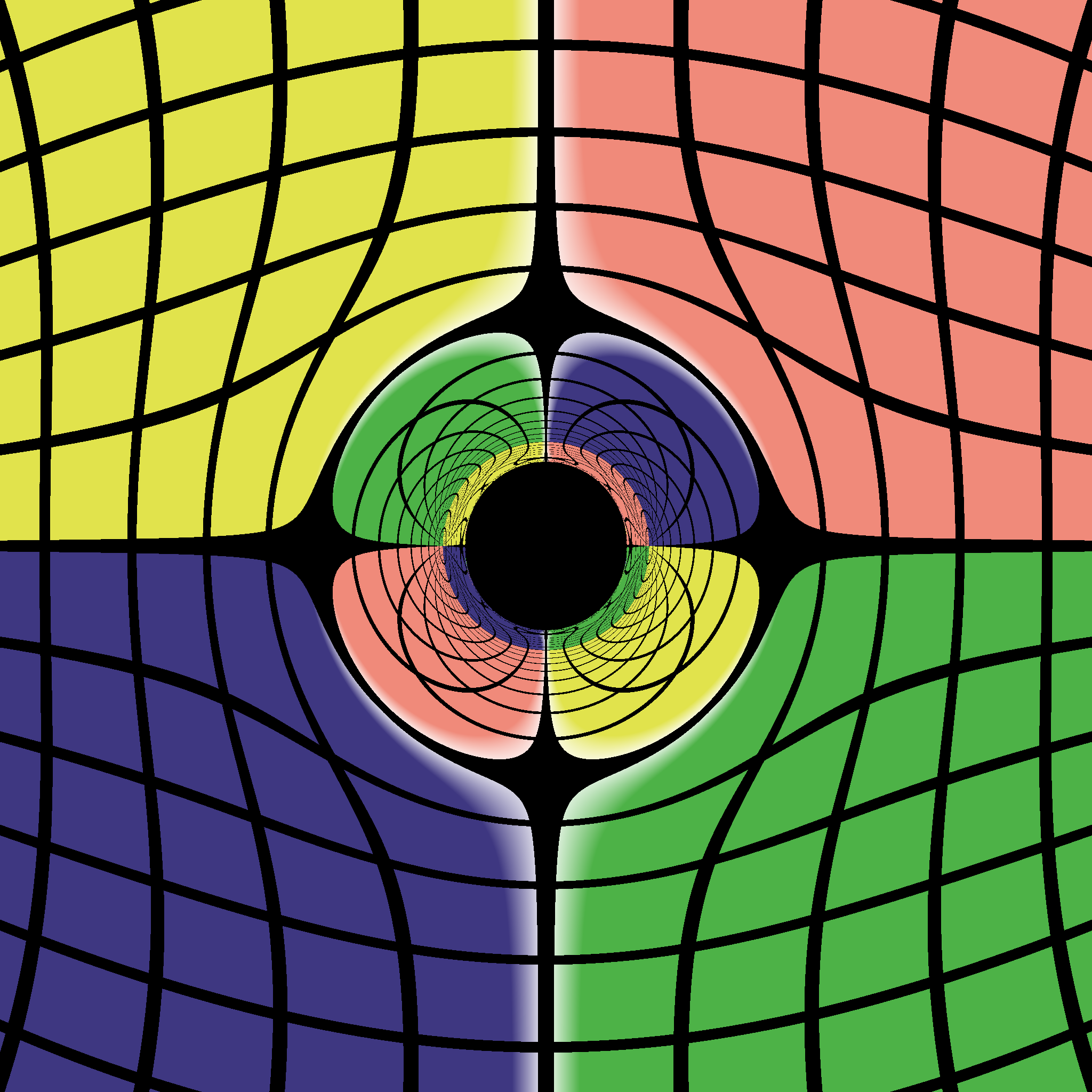}
  \end{minipage}%
  }%
  \subfigure[$R_c=3, \Lambda=-0.66$]{
  \begin{minipage}[t]{0.5\linewidth}
  \centering
  \includegraphics[width=2in]{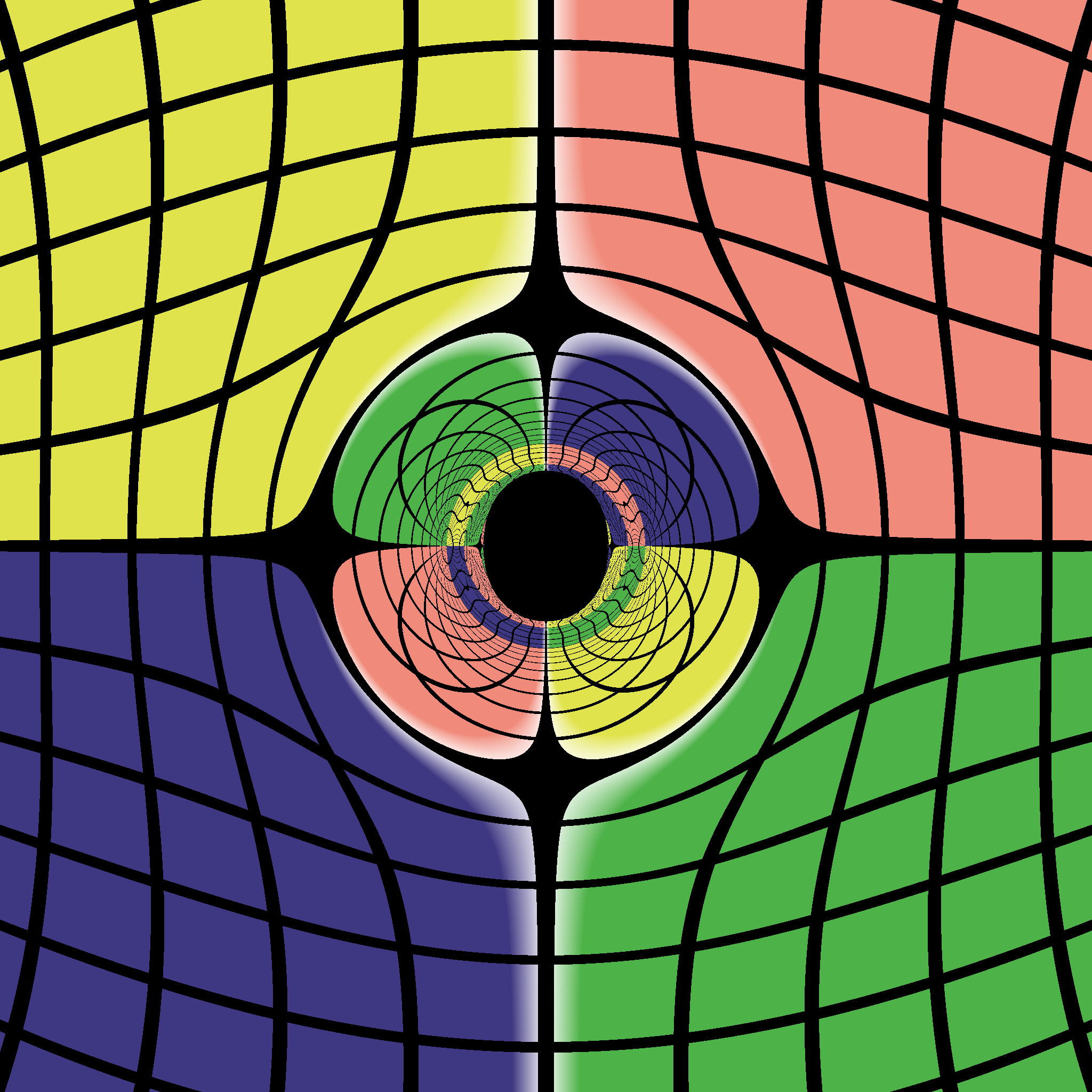}
  \end{minipage}%
  }%

  \subfigure[$R_c=4, \Lambda=-0.165$]{
  \begin{minipage}[t]{0.5\linewidth}
  \centering
  \includegraphics[width=2in]{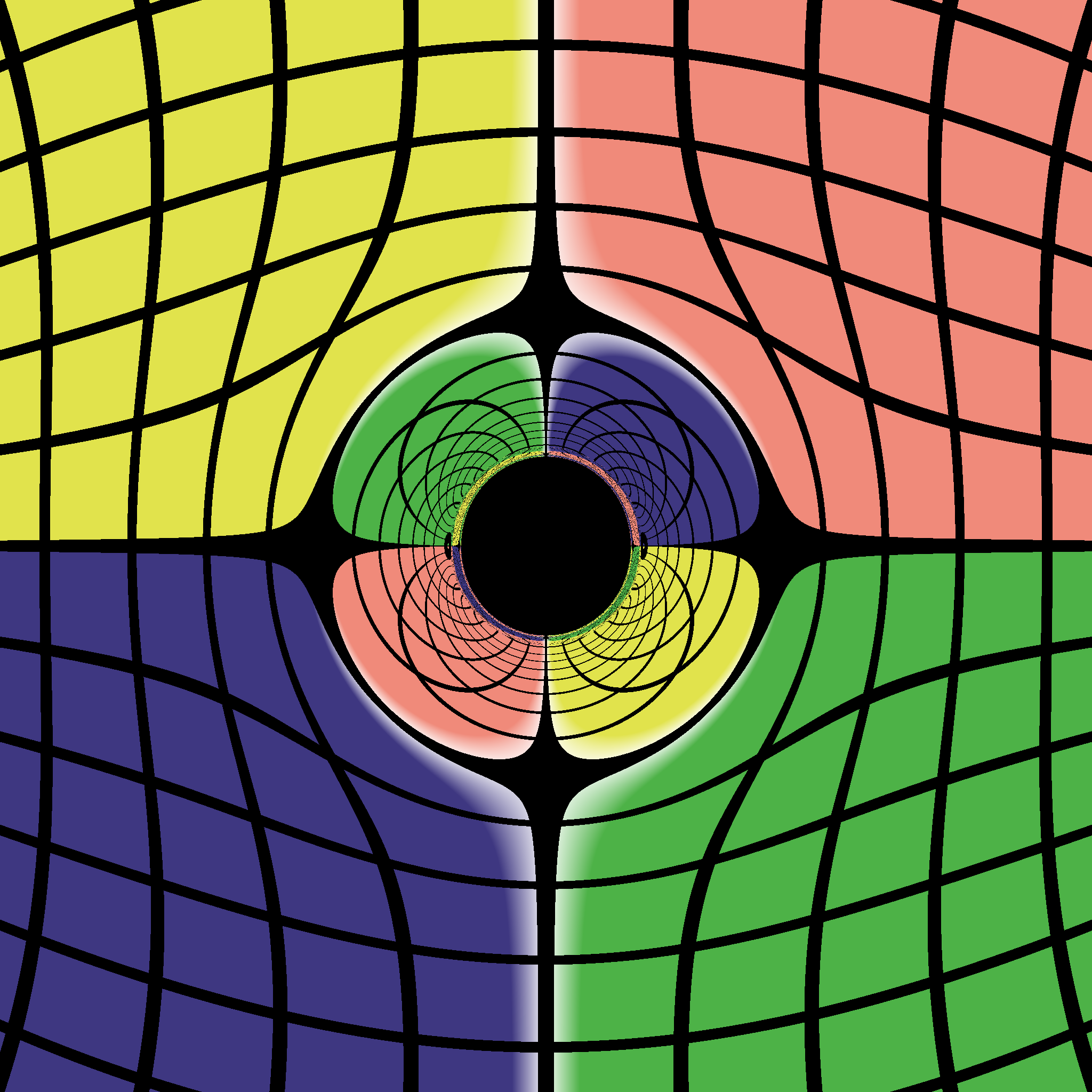}
  \end{minipage}
  }%
  \subfigure[$R_c=4, \Lambda=-0.33$]{
  \begin{minipage}[t]{0.5\linewidth}
  \centering
  \includegraphics[width=2in]{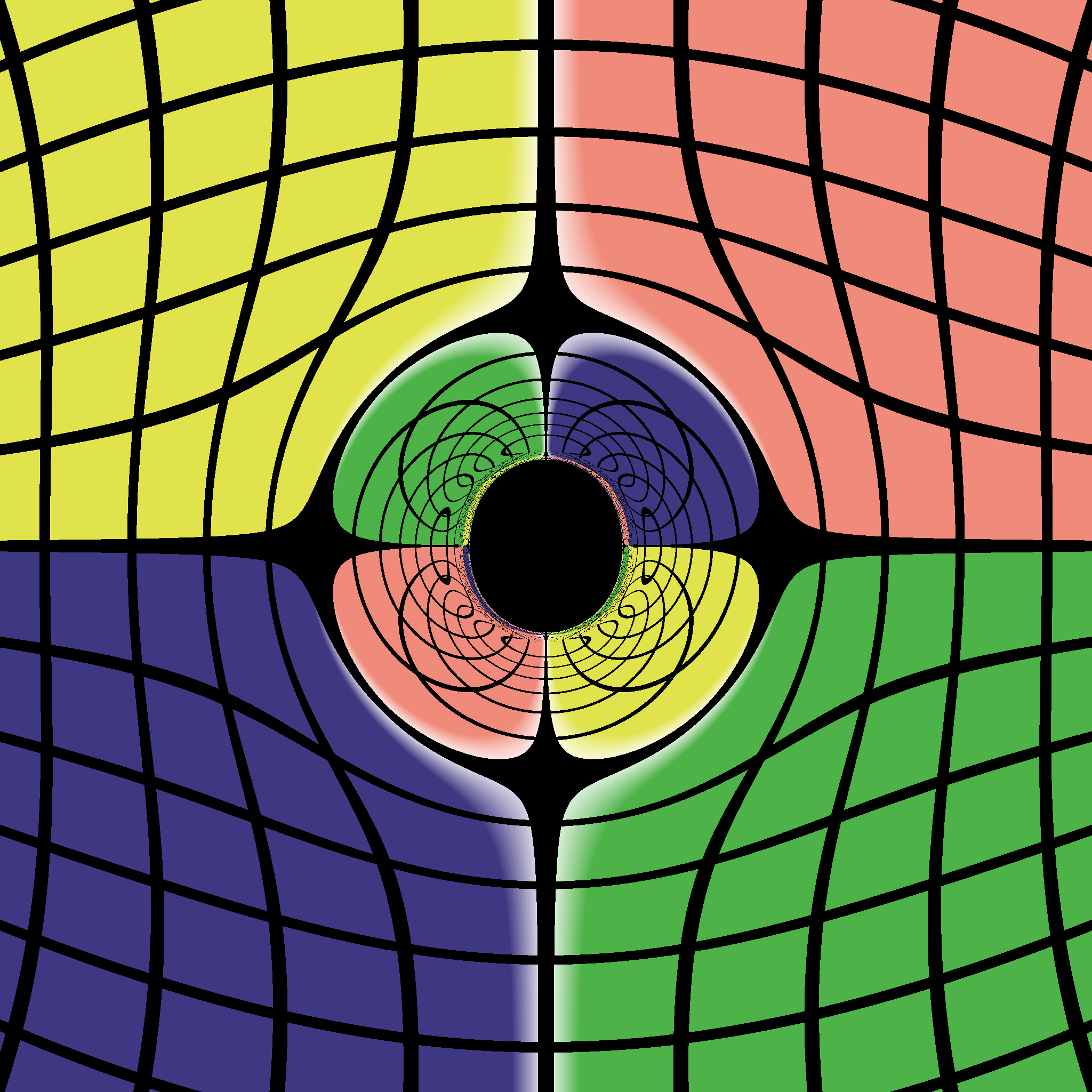}
  \end{minipage}
  }%

  \centering
  \caption{The inclination angle of the observer is fixed at $\theta_o=\pi/2$. $\delta$ is set to be $0.1$.}
  \label{loop}
\end{figure}

Before we move forward to obtain the shadow of the black hole using the numerical backward ray-tracing method, we need to confine the range of $\Lambda$ to make sure the spacetime is well-defined outside the horizon. Because a logarithmic term appears in $V(r)$, we cannot obtain the range of $\Lambda$ analytically. On the other hand, from Eq. (\ref{conti}) , $h(r)$ is obviously a $C^0$ function that will cause discontinuity of Christoffel symbols of the effective metric. As a result, the null geodesics are discontinuous at $R_c$ in terms of the effective spacetime. To overcome this difficulty, we employ a seventh-degree polynomial in a small enough interval $[R_c,R_c+\delta]$ to make $h(r)$ a $C^3$ function \footnote{The reason comes from the fact that the components of the effective metric require $C^2$ and $\partial_r h(r)$ (that is, $Z$ and $W$) appear in the metric}.

Now, we are ready to apply the numerical backward ray-tracing method to this effective metric. As in this case, the position of the current loop, the value of $\Lambda$ and the observing angle all contribute to the shadow of the black hole, it would be very complicated to consider all the conditions. For the readability of the article, we leave the full analysis in further study and list some examples in Fig. \ref{loop}. It appears that  when $|\Lambda|$ becomes large, the deviation from a circle becomes obvious.

\section{Summary}\label{summary}

In this paper, we studied the QED effects on the shadows of the static black holes with magnetic monopoles and the static black holes surrounded with magnetic fields. Firstly, we found that considering QED effect the Hamiltonian give two branches of dispersion relation which correspond to different polarizations of light. As a result, with different polarizations, the observers see black hole shadows of different sizes. This interesting phenomenon may be observed in future EHT observations.

Next, we moved to investigate the shadows of the static black holes with magnetic monopoles. We found an analytical expression of the shadow radius when the effective coupling $|\lambda|\ll1$ or equivalently the black hole mass is large.  In addition, we gave a numerical study considering all the allowed $\lambda$'s, and find that the shadow radius is nearer to the black horizon for smaller black hole due to QED effect.

Then, we turned to consider a more realistic case that a neutral black hole is bathed in magnetic fields. For a Schwarzschild black hole in a uniform magnetic field, we obtained the shadow curves using the numerical backward ray-tracing method. When the observer is located on the equatorial plane, we found that the shadow fitted nicely with an ellipse for small $|\Lambda|$ and the deviation from an ellipse was very small even for a vary large $|\Lambda|$. These facts allowed us to introduce the eccentricity $e$ and discuss  the relation between $e$ and $|\Lambda|$, which turns out to be linear. For the non-equatorial observers, we found the largest deviation from an ellipse occurs at $\theta_o\sim\pi/4$ or $\theta_o\sim3\pi/4$. 
Moreover, we found that the shadow curves for $\theta_o$ and $\pi-\theta_o$ overlap  since the $\mathcal{Z}_2$ symmetry is not broken by the uniform magnetic field. 

Furthermore, we tried to extend our method to a more complicated model in which  there is a current loop around a Schwarzschild black hole. After using some non-trivial numerical techniques, we found our methods worked well for this case and showed some numerical results after fixing $\theta_o=\pi/2$. Considering the fact that many parameters are involved in this model, we leave the complete discussion to the future.

In the present work, we focused on the spherical black holes. It would be interesting to generalize our study to more realistic astrophysical black holes with spinning.

\section*{Acknowledgments}
The work is in part supported by NSFC Grant No. 11335012, No. 11325522, No. 11735001 and No. 11847241. MG and PCL are also supported by NSFC Grant No. 11947210. And MG is also funded by China Postdoctoral Science Foundation Grant No. 2019M660278 and 2020T130020. PCL is also funded by China Postdoctoral Science Foundation Grant No. 2020M670010.

\appendix\label{appen}
\section{Backward ray-tracing method}\label{appendixA}
In this appendix, we would like to give a brief but useful introduction on the numerical backward ray-tracing method which was employed in our study.

Since the effective metric of interest is non-circular and non-rotating in this manuscript, we assume the effective metric takes the form
\bea
ds_{\text{eff}}^2&=&g_{tt}dt^2+g_{rr}dr^2+g_{\theta\theta}d\theta^2+g_{\phi\phi}d\phi^2+X_{rr}dr^2+X_{\theta\theta}d\theta^2+X_{\phi\phi}d\phi^2+2X_{r\theta}drd\theta\nn\\
&&=G_{\mu\nu}dx^\mu dx^\nu
\eea
where $g_{\mu\nu}$ is the metric of the background, that is, Schwarzschild black hole spacetime and $X_{\mu\nu}$ come from the QED effect. The $4$-momentum vector of a photon is written as
\be\label{momenvec}
p^\mu=(\dot{t}, \dot{r}, \dot{\theta}, \dot{\phi})
\ee
where the dot denotes the derivative with respect to the affine parameter $\tau$. By definition, we have the dual vector $p_\mu\equiv g_{\mu\nu}p^\nu$, however, $p^\mu p_\mu\neq0$ in Schwarzschild black hole spacetime. Note that $G_{\mu}p^\mu p^\nu=0$, it is convenient to define a new dual vector
\be
q_\mu\equiv G_{\mu\nu}p^\nu.
\ee
Thus in the effective spacetime, the effective geodesic equations are given by
\bea
\dot{q}_\mu=-\frac{\partial H}{\partial x^\mu},\quad\quad \dot{x}^\mu=\frac{\partial H}{\partial q_\mu}
\eea
in the Hamiltonian canonical formulism, where $H$ is the Hamiltonian of the null particle, $x^\mu$ are  $4$-position of the geodesic. Again, the dot denotes the derivative with respect to the affine parameter $\tau$. Note that these equations are first-order differential equations with the canonical variables $(x^\mu, q_\mu)$, it is very convenient to do the numerical geodesic evolution.

\begin{figure}[t!]
\begin{center}
\includegraphics[width=80mm,angle=0]{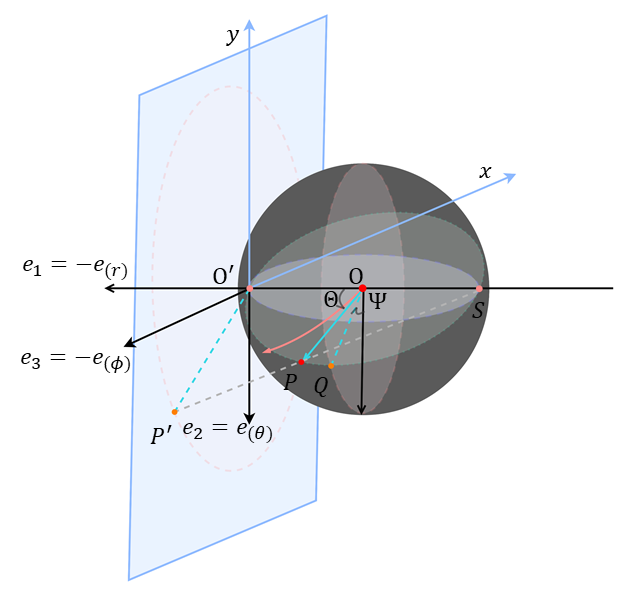}
\end{center}
\caption{In the frame of the observer, the celestial coordinates $\Theta$ and $\Phi$ are introduced to mark each light ray. With the help of the stereographic projection, a useful map is defined from the celestial sphere to the screen of our camera.  }\label{graph1}
\end{figure}

Next, let's skip over explaining how the black hole shadow forms in terms of the null orbits outside the horizon and show the method to determine the boundary curve of the shadow on the observer's sky. We assume an observer is located at $(r_o, \theta_o)$ in the coordinates $\{t, r, \theta, \phi \}$. It is worth mentioning that the QED effect has no effect on the observer, thus in the neighborhood of the observer, a local rest frame can be chosen as
\bea
e_0&=&e_{(t)}=\left.\frac{\partial_t}{\sqrt{-g_{tt}}}\right\vert_{\left(r_o, \theta_o\right)},\nn\\
e_1&=&-e_{(r)}=-\left.\frac{\partial_r}{\sqrt{g_{rr}}}\right\vert_{\left(r_o, \theta_o\right)},\nn\\
e_2&=&e_{(\theta)}=\left.\frac{\partial_\theta}{\sqrt{g_{\theta\theta}}}\right\vert_{\left(r_o, \theta_o\right)},\nn\\
e_3&=&-e_{(\phi)}=-\left.\frac{\partial_\phi}{\sqrt{g_{\phi\phi}}}\right\vert_{\left(r_o, \theta_o\right)}\eea
where $g_{\mu\nu}$ is the metric components of the background. Here, we want to stress that this tetrad is not the only choice, one can choose the right tetrad according to the actual needs. In a real physical process, the light comes to the observer from the source, however, we can consider the light originates from the observer in actual calculation since optical paths are reversible. A diagram is given in Fig. \ref{graph1}, where the observer is located at $O$, the red line represents the light and its arrow indicates the direction of propagation. We use $\overrightarrow{OP}$ to denote the tangent vector of the null geodesic at the point $O$ in the three-dimensional space. To describe the position of the photon seen by the observer, it is convenient to introduce the celestial coordinates as follows. Let's draw a sphere with $O$ as the center and $|\overrightarrow{OP}|$ as the radius, the diameter $O^\prime S$ is parallel to $e_1$ and lies on the equatorial plane, then the line $OO^\prime$ and $OP$ give the angle $\Theta$ which is the first celestial coordinate of the optical image. In addition, the two lines define a plane, combining with the sphere, we can identify the great circle that goes through $P$, $S$ and $O^\prime$. Obviously, this great circle intersects the great circle perpendicular to $e_1$ at two points, the point $Q$ is the one closer to $P$ in Fig. \ref{graph1}. Then vector angle between $\overrightarrow{OQ}$ and $e_2$ is the other celestial coordinate, that is, we have complete celestial coordinates $(\Theta, \Psi)$ hereto. Next, we need move to determine the values of $(\Theta, \Psi)$. On the one hand, in the frame of the observer, the tangent vector of the null geodesic takes in this form
\be\label{framep}
\dot{s}=|\overrightarrow{OP}|\left(-\chi e_0+\cos\Theta e_1+\sin\Theta\cos\Psi e_2+\sin\Theta\sin\Psi e_3\right)
\ee
where the minus sigh guarantees that the tangent vector is pointing to the past. In addition, the undetermined prefactor $\chi$ is introduced to guarantee $p^\mu$ is a null vector in the effective spacetime. Moreover, the components from the QED effect can be expressed as
\be
X_{11}=\frac{X_{rr}}{g_{rr}},\quad X_{22}=\frac{X_{\theta\theta}}{g_{\theta\theta}},\quad X_{12}=X_{21}=-\frac{X_{r\theta}}{\sqrt{g_{rr}g_{\theta\theta}}},\quad X_{11}=\frac{E_{rr}}{g_{rr}},\quad X_{33}=\frac{X_{\phi\phi}}{g_{\phi\phi}},
\ee
in the frame of the observer. Then $p^\mu q_\mu=0$ gives
\be\label{chi}
\chi=\left.\sqrt{1+\frac{X_{rr}}{g_{rr}}\cos^2\Theta-\frac{X_{r\theta}}{\sqrt{g_{rr}g_{\theta\theta}}}\sin2\Theta \cos\Psi+\frac{X_{\theta\theta}}{g_{\theta\theta}}\sin^2\Theta \cos^2\Psi+\frac{X_{\phi\phi}}{g_{\phi\phi}}\sin^2\Theta\sin^2\Psi}\right\vert_{(r_o, \theta_o)}
\ee
In addition, since the photon trajectory is independent of the photon energy, let us introduce the energy of the photon in the frame of the camera and set it to be unity, that is,
\be
1=E_{camera}\equiv G_{ab}p^ae^b_0=|\overrightarrow{OP}|\chi=-\left.\frac{E}{\sqrt{-g_{tt}}}\right\vert_{(r_o, \theta_o)}
\ee
On the other hand, from the Eq. (\ref{momenvec}) we also have the tangent vector expressed in terms of the coordinate bases
\be\label{coorp}
\dot{s}=\dot{t}\partial_t+\dot{r}\partial_r+\dot{\theta}\partial_\theta+\dot{\phi}\partial_\phi.
\ee
where $\dot{}=d/d\tau$. By comparing Eq. (\ref{framep}) and (\ref{coorp}), we conclude that the celestial coordinate of the image is determined, once the values of the 4-momentum of a photon is known. Inversely, if the celestial coordinates of the image are known, the $4$-momentum can be identified. Thus, combing with the position of the observer, the initial values $(x^\mu, q_\mu)$ are determined, then evolve the geodesic equations numerically, we can obtain the path of photons. For some special cases, the explicit expressions exist and the null geodesics are integrable. For example, the null geodesics are integrable in Kerr black hole spacetime since there exists the Carter constant  in addition to the mass, angular momentum and energy. On the contrary, when a spacetime does not possess the Carter constant or one is not able to identify the hidden symmetry, one has to solve the null geodesic equations numerically and determine the component values of 4-the momentum as we mentioned at the beginning of this appendix.

\section{Camera model and stereographic projection}\label{appendixB}

\begin{figure}[t!]
\begin{center}
\includegraphics[width=80mm,angle=0]{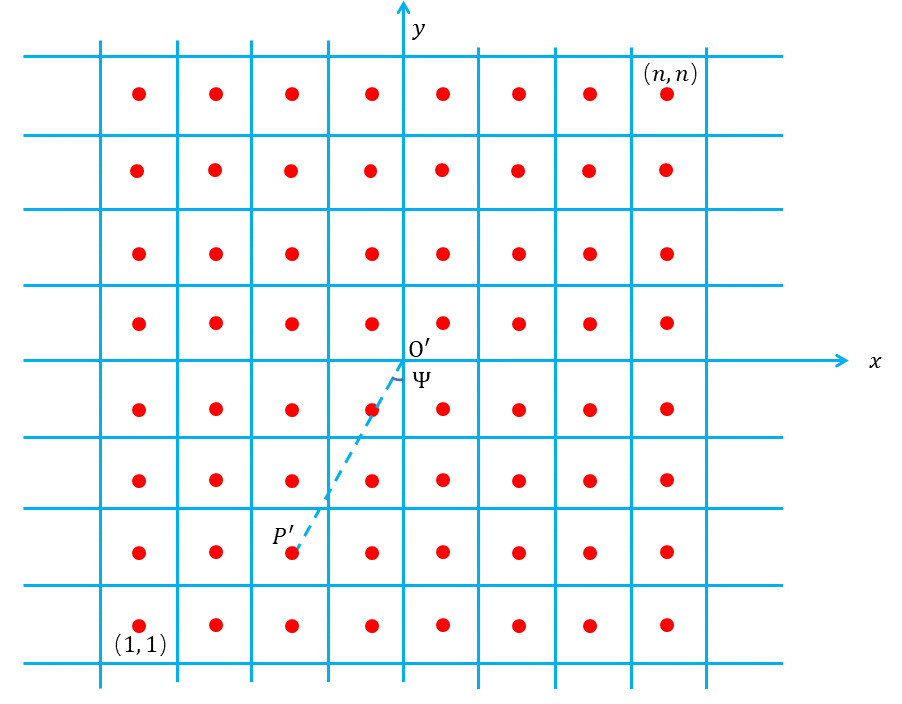}
\end{center}
\caption{An illustration of the camera's screen and pixels, and a standard Cartesian coordinates with the origin $O^\prime$ is set up.}\label{graph2}
\end{figure}

The image of the black hole shadow depends on the camera model. Different models differ in how a viewing direction is mapped onto the plane. As we know, the simplest and most natural one is pinhole camera model with the perspective projection. However, the disadvantage of this model is also obvious that the field of view is relatively narrow. In view of this, we would like to  employ the stereographic projection in this paper, which is often called fisheye camera model.

As shown in Fig. \ref{graph1}, we put the optical centre of our camera at point $S$, then the optical image $P$ on the sphere can be mapped to $P^\prime$ on a plane.  In this plane, we can set up standard Cartesian coordinates with the origin $O^\prime$, shown in Fig. \ref{graph2}. Then the coordinates in this plane are given by
\bea\label{tp}
x_{P^\prime}&=&-2|\overrightarrow{OP}|\tan\frac{\Theta}{2}\sin\Psi\nn\\
y_{P^\prime}&=&-2|\overrightarrow{OP}|\tan\frac{\Theta}{2}\cos\Psi
\eea

\begin{figure}[t!]
\begin{center}
\includegraphics[width=80mm,angle=0]{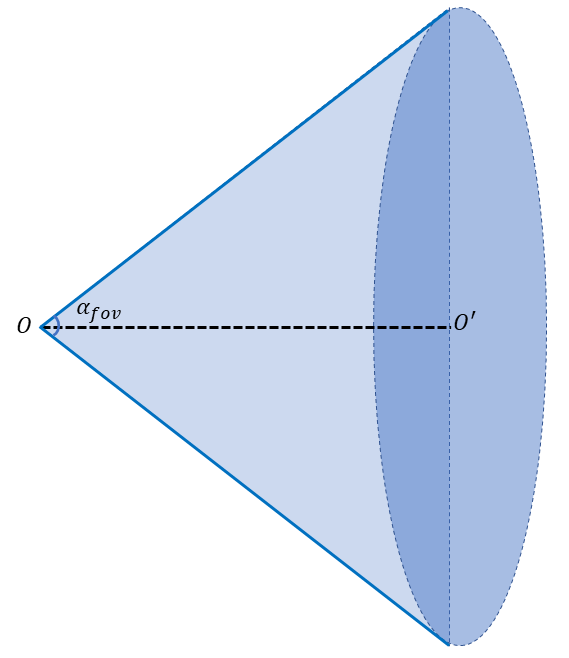}
\end{center}
\caption{The camera's field of view }\label{graph3}
\end{figure}

Next, let us talk about the camera's field of view, which is characterized by the angles in the plane $yO^\prime z$ and plane $xO^\prime z$, respectively. For convenience, we might as well take two equal angles. An illustrated field of view is given in Fig. \ref{graph3}. Then, the length of the square screen is given by
\be
L=2|\overrightarrow{OP}|\tan\frac{\al}{2}
\ee
Let's suppose we take $n\times n$ pixels on the screen, the length occupied per pixel is
\be
\ell=\frac{2|\overrightarrow{OP}|}{n}\tan\frac{\al}{2}
\ee
We mark the pixels with $(i, j)$, and set the pixel on the bottom left corner (the top right corner) to $(1, 1)$ ($(n, n)$). Also, $i$ and $j$ are chosen from $1$ to $n$. Combining the cartesian coordinates of the center point of a pixel, we have
\bea\label{ij}
x_{P^\prime}&=&\ell\left(i-\frac{n+1}{2}\right)\nn\\
y_{P^\prime}&=&\ell\left(j-\frac{n+1}{2}\right)
\eea
Comparing Eq. (\ref{tp}) with (\ref{ij}), we can establish a relation between $(i, j)$ and $(\Theta, \Psi)$ as follows
\bea
\tan\Psi&=&\frac{j-(n+1)/2}{i-(n+1)/2}\nn\\
\tan\frac{\Theta}{2}&=&\tan\frac{\al}{2}\frac{\sqrt{\left[i-(n+1)/2\right]^2+\left[j-(n+1)/2\right]^2}}{n}
\eea

\section{Source}\label{appendixC}

\begin{figure}[t!]
\begin{center}
\includegraphics[width=80mm,angle=0]{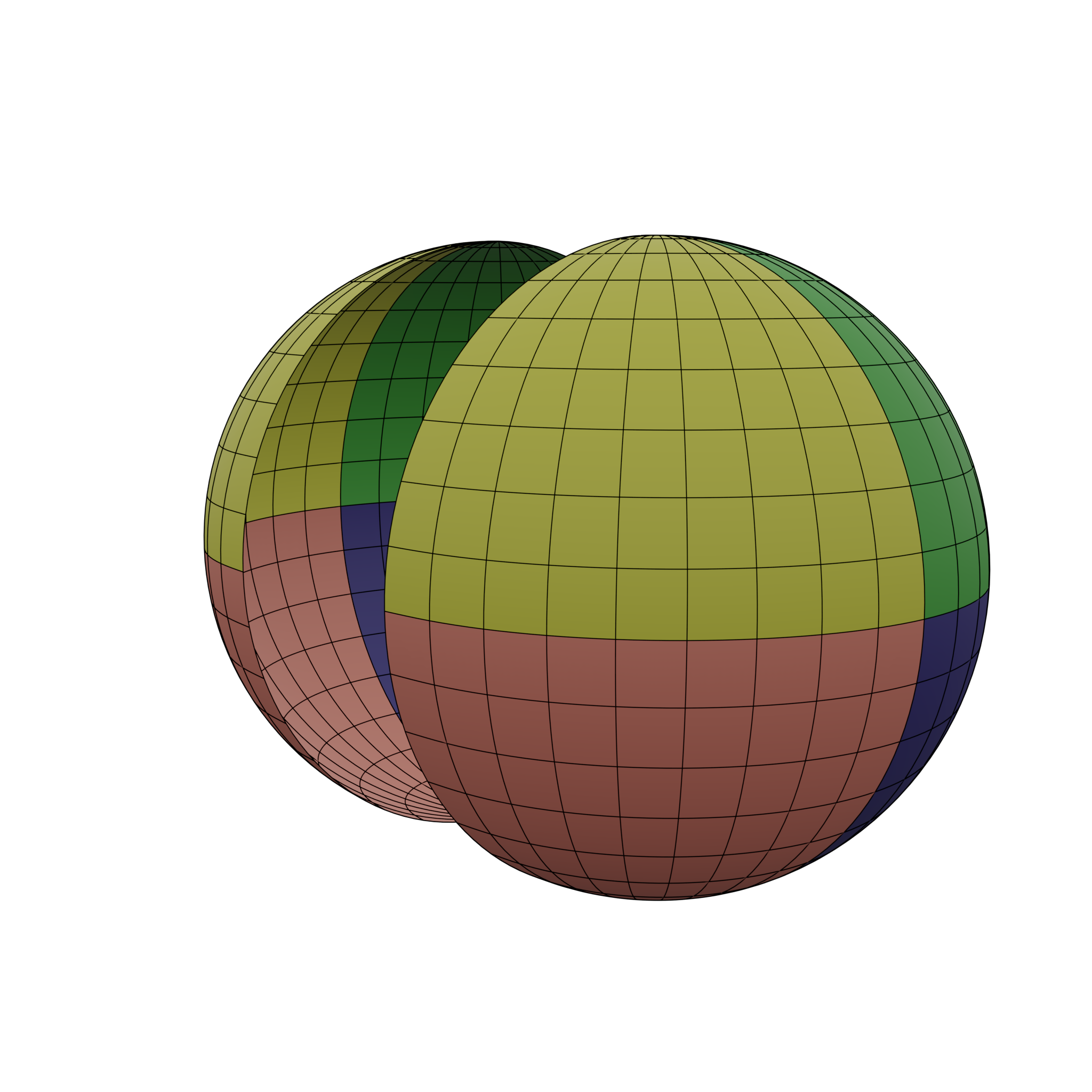}
\end{center}
\caption{An illustration of our spherical light source at infinity.}\label{sphere}
\end{figure}
In our work, we use the extended source to illuminate the system. Our model can be seen in Fig.\ref{sphere} where we cut the photosphere to reveal its interior. We divide the ball into grids with the network made of latitude and longitude lines. The interval for two adjacent longitude or latitude line is $\pi/18$. To create the image of the black hole, we need to assign an appropriate single color to each grid. Here, we divided the extended source into four colors. Then the color of the pixels on the screen can be determined based on the corresponding source point when these null geodesics originate from the extended source. And the dark area means the corresponding photons are captured by the black hole.


\begin{thebibliography}{10}


\bibitem{Akiyama:2019cqa}
  K.~Akiyama {\it et al.} [Event Horizon Telescope Collaboration],
  Astrophys.\ J.\  {\bf 875}, no. 1, L1 (2019).

 \bibitem{Akiyama:2019brx}
K.~Akiyama {\it et al.} [Event Horizon Telescope Collaboration],
  Astrophys.\ J.\  {\bf 875}, no. 1, L2 (2019).

  \bibitem{Akiyama:2019sww}
  K.~Akiyama {\it et al.} [Event Horizon Telescope Collaboration],
  Astrophys.\ J.\  {\bf 875}, no. 1, L3 (2019).

  \bibitem{Akiyama:2019bqs}
  K.~Akiyama {\it et al.} [Event Horizon Telescope Collaboration],
  Astrophys.\ J.\  {\bf 875}, no. 1, L4 (2019).

  \bibitem{Akiyama:2019fyp}
  K.~Akiyama {\it et al.} [Event Horizon Telescope Collaboration],
  Astrophys.\ J.\  {\bf 875}, no. 1, L5 (2019).

  \bibitem{Akiyama:2019eap}
  K.~Akiyama {\it et al.} [Event Horizon Telescope Collaboration],
  Astrophys.\ J.\  {\bf 875}, no. 1, L6 (2019).

  \bibitem{Takahashi:2005hy}
R.~Takahashi,
Publ. Astron. Soc. Jap. \textbf{57} (2005), 273.

\bibitem{Hioki:2009na}
K.~Hioki and K.~i.~Maeda,
Phys. Rev. D \textbf{80} (2009), 024042.

\bibitem{Tsupko:2017rdo}
O.~Y.~Tsupko,
Phys. Rev. D \textbf{95} (2017) no.10, 104058.

\bibitem{Kumar:2018ple}
R.~Kumar and S.~G.~Ghosh,
Astrophys. J. \textbf{892} (2020), 78.

\bibitem{Tamburini:2019vrf}
F.~Tamburini, B.~Thid\'e and M.~Della Valle,
Mon. Not. Roy. Astron. Soc. \textbf{492} (2020) no.1, L22-L27.

  \bibitem{Bambi:2019tjh}
C.~Bambi, K.~Freese, S.~Vagnozzi and L.~Visinelli,
Phys. Rev. D \textbf{100} (2019) no.4, 044057.

\bibitem{Dokuchaev:2019pcx}
V.~I.~Dokuchaev and N.~O.~Nazarova,
Universe \textbf{5} (2019), 183.

\bibitem{Gralla:2020yvo}
S.~E.~Gralla and A.~Lupsasca,
[arXiv:2007.10336 [gr-qc]].

\bibitem{Gralla:2020srx}
S.~E.~Gralla, A.~Lupsasca and D.~P.~Marrone,
[arXiv:2008.03879 [gr-qc]].

\bibitem{Shaikh:2018lcc}
R.~Shaikh, P.~Kocherlakota, R.~Narayan and P.~S.~Joshi,
Mon. Not. Roy. Astron. Soc. \textbf{482} (2019) no.1, 52-64.

\bibitem{Dai:2019mse}
D.~C.~Dai and D.~Stojkovic,
Phys. Rev. D \textbf{100} (2019) no.8, 083513.

\bibitem{Joshi:2020tlq}
A.~B.~Joshi, D.~Dey, P.~S.~Joshi and P.~Bambhaniya,
Phys. Rev. D \textbf{102} (2020) no.2, 024022.

\bibitem{Wang:2020emr}
X.~Wang, P.~C.~Li, C.~Y.~Zhang and M.~Guo,
Phys. Lett. B \textbf{811} (2020), 135930.

\bibitem{Paul:2020ufc}
S.~Paul,
Phys. Rev. D \textbf{102} (2020) no.6, 064045.

\bibitem{Dey:2020bgo}
D.~Dey, R.~Shaikh and P.~S.~Joshi,
[arXiv:2009.07487 [gr-qc]].

\bibitem{Wielgus:2020uqz}
M.~Wielgus, J.~Horak, F.~Vincent and M.~Abramowicz,
Phys. Rev. D \textbf{102} (2020) no.8, 084044.

\bibitem{Zhang:2020xub}
M.~Zhang and J.~Jiang,
[arXiv:2010.12194 [gr-qc]].

\bibitem{Bambi:2008jg}
C.~Bambi and K.~Freese,
Phys. Rev. D \textbf{79} (2009), 043002.

\bibitem{Johannsen:2010ru}
T.~Johannsen and D.~Psaltis,
Astrophys. J. \textbf{718} (2010), 446-454.

\bibitem{Amarilla:2011fx}
L.~Amarilla and E.~F.~Eiroa,
Phys. Rev. D \textbf{85} (2012), 064019.

\bibitem{Grenzebach:2015oea}
  A.~Grenzebach, V.~Perlick and C.~Lämmerzahl,
  Int.\ J.\ Mod.\ Phys.\ D {\bf 24}, no. 09, 1542024 (2015).

\bibitem{Amir:2016cen}
  M.~Amir and S.~G.~Ghosh,
  Phys.\ Rev.\ D {\bf 94}, no. 2, 024054 (2016).

  \bibitem{Abdujabbarov:2016hnw}
  A.~Abdujabbarov, M.~Amir, B.~Ahmedov and S.~G.~Ghosh,
  Phys.\ Rev.\ D {\bf 93}, no. 10, 104004 (2016).


  \bibitem{Dastan:2016vhb}
  S.~Dastan, R.~Saffari and S.~Soroushfar,
  arXiv:1606.06994 [gr-qc].

\bibitem{Wang:2017hjl}
  M.~Wang, S.~Chen and J.~Jing,
  JCAP {\bf 1710}, 051 (2017).

   \bibitem{Cunha:2018acu}
  P.~V.~P.~Cunha and C.~A.~R.~Herdeiro,
  Gen.\ Rel.\ Grav.\  {\bf 50}, no. 4, 42 (2018).

  \bibitem{Wang:2018eui}
  M.~Wang, S.~Chen and J.~Jing,
  Phys.\ Rev.\ D {\bf 98}, no. 10, 104040 (2018).

  \bibitem{Hennigar:2018hza}
  R.~A.~Hennigar, M.~B.~J.~Poshteh and R.~B.~Mann,
  Phys.\ Rev.\ D {\bf 97}, no. 6, 064041 (2018).

\bibitem{Perlick:2018iye}
  V.~Perlick, O.~Y.~Tsupko and G.~S.~Bisnovatyi-Kogan,
  Phys.\ Rev.\ D {\bf 97}, no. 10, 104062 (2018).

\bibitem{Bisnovatyi-Kogan:2018vxl}
  G.~S.~Bisnovatyi-Kogan and O.~Y.~Tsupko,
  Phys.\ Rev.\ D {\bf 98}, no. 8, 084020 (2018).

   \bibitem{Vagnozzi:2019apd}
S.~Vagnozzi and L.~Visinelli,
Phys. Rev. D \textbf{100} (2019) no.2, 024020.

  \bibitem{Banerjee:2019nnj}
I.~Banerjee, S.~Chakraborty and S.~SenGupta,
Phys. Rev. D \textbf{101} (2020) no.4, 041301.

\bibitem{Tsupko:2019mfo}
  O.~Y.~Tsupko and G.~S.~Bisnovatyi-Kogan,
 Int. J. Mod. Phys. D \textbf{29} (2020) no.09, 2050062.

  \bibitem{Ovgun:2018tua}
  A.~Övgün, İ.~Sakalli and J.~Saavedra,
  JCAP {\bf 1810}, 041 (2018).

  \bibitem{Wang:2018prk}
  H.~M.~Wang, Y.~M.~Xu and S.~W.~Wei,
  JCAP {\bf 1903}, 046 (2019).

  \bibitem{Wei:2019pjf}
  S.~W.~Wei, Y.~C.~Zou, Y.~X.~Liu and R.~B.~Mann,
  JCAP {\bf 1908}, 030 (2019).

  \bibitem{Kumar:2019ohr}
  R.~Kumar, B.~P.~Singh and S.~G.~Ghosh,
  Annals Phys. \textbf{420} (2020), 168252.

  \bibitem{Shaikh:2019fpu}
  R.~Shaikh,
  Phys.\ Rev.\ D {\bf 100}, no. 2, 024028 (2019).

  \bibitem{Contreras:2019nih}
  E.~Contreras, J.~M.~Ramirez-Velasquez, Á.~Rincón, G.~Panotopoulos and P.~Bargueño,
  Eur.\ Phys.\ J.\ C {\bf 79}, no. 9, 802 (2019).

  \bibitem{Ovgun:2019jdo}
  A.~Övgün, İ.~Sakalli, J.~Saavedra and C.~Leiva,
  Mod. Phys. Lett. A \textbf{35} (2020) no.20, 2050163.

  \bibitem{Contreras:2019cmf}
  E.~Contreras, Á.~Rincón, G.~Panotopoulos, P.~Bargueño and B.~Koch,
  Phys. Rev. D \textbf{101} (2020) no.6, 064053.

\bibitem{Wang:2019skw}
M.~Wang, S.~Chen and J.~Jing,
[arXiv:1908.04527 [gr-qc]].

  \bibitem{Das:2019sty}
  A.~Das, A.~Saha and S.~Gangopadhyay,
  Eur. Phys. J. C \textbf{80} (2020) no.3, 180.

\bibitem{Zhang:2019glo}
M.~Zhang and M.~Guo,
Eur. Phys. J. C \textbf{80} (2020) no.8, 790.

  \bibitem{Lu:2019zxb}
  H.~Lu and H.~D.~Lyu,
  Phys. Rev. D \textbf{101} (2020) no.4, 044059.

\bibitem{Dokuchaev:2019jqq}
V.~I.~Dokuchaev and N.~O.~Nazarova,
Phys. Usp. \textbf{63} (2020), 583.

  \bibitem{Feng:2019zzn}
  X.~H.~Feng and H.~Lu,
  Eur. Phys. J. C \textbf{80} (2020) no.6, 551.

  \bibitem{Kumar:2019pjp}
  R.~Kumar, S.~G.~Ghosh and A.~Wang,
  Phys.\ Rev.\ D {\bf 100}, no. 12, 124024 (2019).

  \bibitem{Ma:2019ybz}
  L.~Ma and H.~Lu,
  Phys. Lett. B \textbf{807} (2020), 135535.

  \bibitem{Kumar:2020hgm}
  R.~Kumar, S.~G.~Ghosh and A.~Wang,
  Phys. Rev. D \textbf{101} (2020) no.10, 104001.


\bibitem{Li:2020drn}
P.~C.~Li, M.~Guo and B.~Chen,
Phys. Rev. D \textbf{101} (2020) no.8, 084041.

\bibitem{Chang:2020miq}
Z.~Chang and Q.~H.~Zhu,
Phys. Rev. D \textbf{101} (2020) no.8, 084029.

\bibitem{Wei:2020ght}
S.~W.~Wei and Y.~X.~Liu,
[arXiv:2003.07769 [gr-qc]].

\bibitem{Pantig:2020uhp}
R.~C.~Pantig and E.~T.~Rodulfo,
Chin. J. Phys. \textbf{68} (2020), 236-257.

\bibitem{Roy:2020dyy}
R.~Roy and S.~Chakrabarti,
Phys. Rev. D \textbf{102} (2020) no.2, 024059.

\bibitem{Guo:2020zmf}
M.~Guo and P.~C.~Li,
Eur. Phys. J. C \textbf{80} (2020) no.6, 588.

\bibitem{Chen:2020aix}
C.~Y.~Chen,
JCAP \textbf{05} (2020), 040.

\bibitem{EslamPanah:2020hoj}
B.~Eslam Panah, K.~Jafarzade and S.~H.~Hendi,
[arXiv:2004.04058 [hep-th]].

\bibitem{Zeng:2020dco}
X.~X.~Zeng, H.~Q.~Zhang and H.~Zhang,
Eur. Phys. J. C \textbf{80} (2020) no.9, 872.

\bibitem{Ovgun:2020gjz}
A.~\"Ovg\"un and \.I.~Sakalli, Class. Quant. Grav. \textbf{37} (2020) no.22, 225003.

\bibitem{Jusufi:2020cpn}
K.~Jusufi, M.~Jamil and T.~Zhu,
Eur. Phys. J. C \textbf{80} (2020) no.5, 354.

\bibitem{Badia:2020pnh}
J.~Badia and E.~F.~Eiroa,
Phys. Rev. D \textbf{102} (2020) no.2, 024066.

\bibitem{Belhaj:2020nqy}
A.~Belhaj, L.~Chakhchi, H.~El Moumni, J.~Khalloufi and K.~Masmar,
Int. J. Mod. Phys. A \textbf{35} (2020) no.27, 2050170.

\bibitem{Khodadi:2020jij}
M.~Khodadi, A.~Allahyari, S.~Vagnozzi and D.~F.~Mota,
JCAP \textbf{09} (2020), 026.

\bibitem{Allahyari:2019jqz}
A.~Allahyari, M.~Khodadi, S.~Vagnozzi and D.~F.~Mota,
JCAP \textbf{02}, 003 (2020)

\bibitem{Khodadi:2020gns}
M.~Khodadi and E.~N.~Saridakis,
[arXiv:2012.05186 [gr-qc]].


\bibitem{Ghosh:2020tdu}
D.~Ghosh, A.~Thalapillil and F.~Ullah,
[arXiv:2009.03363 [hep-ph]].


\bibitem{Chang:2020lmg}
Z.~Chang and Q.~H.~Zhu,
Phys. Rev. D \textbf{102} (2020) no.4, 044012.

\bibitem{Belhaj:2020rdb}
A.~Belhaj, M.~Benali, A.~El Balali, H.~El Moumni and S.~E.~Ennadifi,
Class. Quant. Grav. \textbf{37} (2020) no.21, 215004.

\bibitem{Zeng:2020vsj}
X.~X.~Zeng and H.~Q.~Zhang,
Eur. Phys. J. C \textbf{80} no.11, 1058.

\bibitem{Dokuchaev:2020wqk}
V.~I.~Dokuchaev and N.~O.~Nazarova,
Universe \textbf{6} (2020) no.9, 154.

\bibitem{Peng:2020wun}
J.~Peng, M.~Guo and X.~H.~Feng,
[arXiv:2008.00657 [gr-qc]].

\bibitem{Belhaj:2020mlv}
A.~Belhaj, M.~Benali, A.~E.~Balali, W.~E.~Hadri, H.~El Moumni and E.~Torrente-Lujan,
[arXiv:2008.09908 [hep-th]].

\bibitem{Belhaj:2020okh}
A.~Belhaj, H.~Belmahi, M.~Benali, W.~E.~Hadri, H.~El Moumni and E.~Torrente-Lujan,
[arXiv:2008.13478 [hep-th]].

\bibitem{Psaltis:2020lvx}
D.~Psaltis \textit{et al.} [Event Horizon Telescope],
Phys. Rev. Lett. \textbf{125} (2020) no.14, 141104.

\bibitem{Contreras:2020kgy}
E.~Contreras, \'A.~Rinc\'on, G.~Panotopoulos and P.~Bargue\~no,
[arXiv:2010.03734 [gr-qc]].

\bibitem{Gralla:2020pra}
S.~E.~Gralla,
[arXiv:2010.08557 [astro-ph.HE]].

\bibitem{Cotaescu:2020kcr}
I.~I.~Cotaescu,
[arXiv:2011.02434 [gr-qc]].

\bibitem{Volkel:2020xlc}
S.~H.~V\"olkel, E.~Barausse, N.~Franchini and A.~E.~Broderick,
[arXiv:2011.06812 [gr-qc]].

\bibitem{Guo:2018kis}
M.~Guo, N.~A.~Obers and H.~Yan,
Phys. Rev. D \textbf{98} (2018) no.8, 084063.

\bibitem{Yan:2019etp}
H.~Yan,
Phys. Rev. D \textbf{99} (2019) no.8, 084050.

\bibitem{Guo:2019pte}
M.~Guo, P.~C.~Li and B.~Chen,
Phys. Rev. D \textbf{101} (2020) no.2, 024054.

\bibitem{Guo:2019lur}
M.~Guo, S.~Song and H.~Yan,
Phys. Rev. D \textbf{101} (2020) no.2, 024055.

\bibitem{Li:2020val}
P.~C.~Li, M.~Guo and B.~Chen,
[arXiv:2006.05153 [gr-qc]].

\bibitem{Guo:2020qwk}
M.~Guo and S.~Gao,
[arXiv:2011.02211 [gr-qc]].

\bibitem{Abramowicz:2011xu}
M.~A.~Abramowicz and P.~C.~Fragile,
Living Rev. Rel. \textbf{16} (2013), 1.


\bibitem{Eatough:2013nva}
R.~P.~Eatough, H.~Falcke, R.~Karuppusamy, K.~J.~Lee, D.~J.~Champion, E.~F.~Keane, G.~Desvignes, D.~H.~F.~M.~Schnitzeler, L.~G.~Spitler and M.~Kramer, \textit{et al.}
Nature \textbf{501} (2013), 391-394.

\bibitem{Doeleman:2012zc}
S.~S.~Doeleman, V.~L.~Fish, D.~E.~Schenck, C.~Beaudoin, R.~Blundell, G.~C.~Bower, A.~E.~Broderick, R.~Chamberlin, R.~Freund and P.~Friberg, \textit{et al.}
Science \textbf{338} (2012), 355.

\bibitem{DeLorenci:2000yh}
V.~De Lorenci, R.~Klippert, M.~Novello and J.~Salim,
Phys. Lett. B \textbf{482}, no.1-3, 134-140 (2000).

  \bibitem{Novello:1999pg}
M.~Novello, V.~De Lorenci, J.~Salim and R.~Klippert,
Phys. Rev. D \textbf{61}, 045001 (2000).

\bibitem{McDonald:2019wou}
J.~I.~McDonald and L.~B.~Ventura,
Phys. Rev. D \textbf{101} (2020) no.12, 123503.

\bibitem{Schwarz:2020jjh}
D.~J.~Schwarz, J.~Goswami and A.~Basu,
[arXiv:2003.10205 [hep-ph]].

\bibitem{Chen:2020qyp}
S.~Chen, M.~Wang and J.~Jing,
JHEP \textbf{07} (2020), 054.


\bibitem{Heisenberg:1935qt}
  W.~Heisenberg and H.~Euler,
  Z.\ Phys.\  {\bf 98}, no. 11-12, 714 (1936).


\bibitem{Drummond:1979pp}
I.~T.~Drummond and S.~J.~Hathrell,
Phys. Rev. D \textbf{22} (1980), 343.

\bibitem{Ahmadiniaz:2020lbg}
N.~Ahmadiniaz, T.~E.~Cowan, R.~Sauerbrey, U.~Schramm, H.~P.~Schlenvoigt and R.~Sch\"utzhold,
Phys. Rev. D \textbf{101} (2020) no.11, 116019.


\bibitem{Karbstein:2016hlj}
F.~Karbstein,
[arXiv:1611.09883 [hep-th]].

\bibitem{Karbstein:2019oej}
F.~Karbstein,
Particles \textbf{3}, no.1, 39-61 (2020)



\bibitem{Zhong2020}
Z.~Zhong, Z.~Hu, M.~Guo, and B. Chen, Work in progress.

\bibitem{Wald:1974np}
  R.~M.~Wald,
  Phys.\ Rev.\ D {\bf 10}, 1680 (1974).

\bibitem{Carter1972}
B. Carter, in Black- Holes Les Houches 1972, edited by C. DeWitt and B. S. DeWitt (Gordon and Breach, New York, 1973).

\bibitem{Petterson:1974bt}
J.~A.~Petterson,
Phys. Rev. D \textbf{10} (1974), 3166-3170.

\end{thebibliography}
\end{document}